\documentclass[useAMS,onecolumn]{mn2e}

\newif\ifAMStwofonts

\usepackage{subfigure}
\usepackage[dvips]{graphicx}
\usepackage{amssymb}	
\usepackage{epsfig}
\usepackage{color}

\def\pmb#1{\mbox{\boldmath$#1$}}

\def\ltsim {\lesssim}

\def\be{\begin{equation}}
\def\ee{\end{equation}}

\def\pmbmt#1{\pmb{\sf #1}}
\def\rmi{{\rm i}}

\begin{document}

\title[Axisymmetric modes of magnetized stars]{Axisymmetric magnetic modes of neutron stars having mixed poloidal and toroidal magnetic fields}

\author[U. Lee]{
Umin Lee$^{1}$\thanks{E-mail: lee@astr.tohoku.ac.jp},
\\
$^{1}$Astronomical Institute, Tohoku University, Sendai, Miyagi 980-8578, Japan\\
}

\date{Accepted XXX. Received YYY; in original form ZZZ}
\pubyear{2017}

\maketitle

\begin{abstract}
We calculate axisymmetric magnetic modes of a neutron star possessing a mixed poloidal and toroidal magnetic field, where the toroidal field is assumed to be proportional to a dimensionless parameter $\zeta_0$.
Here, we assume an isentropic structure for the neutron star and consider no effects of rotation.
Ignoring the equilibrium deformation due to the magnetic field, we employ a polytrope of the index $n=1$ as the background model for our modal analyses.
For the mixed poloidal and toroidal magnetic field with $\zeta_0\not=0$, 
axisymmetric spheroidal and toroidal modes are coupled.
We compute axisymmetric spheroidal and toroidal magnetic modes as a function of the parameter $\zeta_0$ from $0$ to $\sim 1$ for the surface field strengths $B_S=10^{14}$G and $10^{15}$G.
We find that the frequency $\omega$ of the magnetic modes decreases with increasing $\zeta_0$.
We also find that the frequency of the spheroidal magnetic modes is almost exactly proportional to $B_S$ for $\zeta_0\lesssim 1$ but that this proportionality holds only when $\zeta_0\ll 1$ for the toroidal magnetic modes.
The wave patterns of the spheroidal magnetic modes and toroidal magnetic modes are not strongly 
affected by the coupling so long as $\zeta_0\lesssim 1$.
We find no unstable modes having $\omega^2<0$.
\end{abstract}


\section{Introduction}

Quasi-periodic-oscillations (QPOs) found in the tail of the giant X/$\gamma$-ray flares of SGR 1806-204 (Israel et al 2005)  and SGR 1900+14 (e.g., Strohmayer \& Watts 2005, 2006, Watts \& Strohmayer 2006) have brought about an intense interest in the
oscillations of magnetars, i.e., strongly magnetized neutron stars (e.g., Woods \& Thompson 2006; Mereghetti 2008).
Because of the suggestion made by Duncan (1998) before the detection of the magnetar QPOs,
the crustal torsional oscillations of magnetized neutron stars 
were first investigated as a candidate for
the QPOs (e.g., Glampedakis et al 2006).
However, motivated by the suggestion that the torsional modes in the solid crust will be quickly damped by frequency resonance with Alfv\'en continuum in the fluid core (Levin 2006, 2007), 
many researches carried out MHD simulations that follow time evolution of small amplitude and global toroidal perturbations to investigate modal properties of magnetars 
(e.g., Sotani et al. 2008; Cerd\'a-Dur\'an et al. 2009; Colaiuda \& Kokkotas 2011; Gabler et al. 2011, 2012).
For example, Gabler et al (2011, 2012) showed that because of resonant damping associated
with Alfv\'en continuum in the fluid core the crustal modes cannot survive long enough to explain the observed QPOs.
They also suggested that Alfv\'en modes, instead of crustal modes, could be responsible for the QPOs if the field strength is higher than $\sim 10^{15}$G since the damping timescales of Alfv\'en modes
are much longer than that of the crustal torsional modes suffering resonant damping with Alfv\'en continuum.

These early studies of the oscillations of magnetars are mainly concerned with axisymmetric toroidal modes
of the stars, expecting that toroidal modes are probably easily excited to observable amplitudes
since they produce no density perturbations.
For axisymmetric oscillations of magnetized stars, spheroidal and toroidal components of
the perturbed velocity fields are decoupled
and can be treated separately for a purely poloidal or toroidal magnetic field configuration for
non-rotating stars.
This property remains true even for general relativistic treatment.
For non-axisymmetric oscillations of magnetized stars, however, the toroidal 
velocity component is coupled with the spheroidal one which accompanies the density and pressure perturbations,
even if we assume a pure poloidal or toroidal magnetic field for non-rotating stars.

Lander et al. (2010), Passamonti \& Lander (2013) discussed, using MHD simulations, 
such non-axisymmetric oscillation modes of magnetized stars assuming a purely toroidal magnetic field.
For rotating magnetized stars, the modal property will be more complicated because the spheroidal (polar) and toroidal (axial) components of the perturbed velocity fields are coupled and rotational modes such as
inertial modes and $r$-modes come in as additional classes of oscillation modes.
Lander \& Jones (2011) calculated non-axisymmetric oscillations of magnetized rotating stars for
a purely poloidal magnetic field. 
They obtained polar-led Alfv\'en modes which reduce to inertial modes in the limit of ${\cal M}/{\cal T}\to 0$, where $\cal M$ and $\cal T$ are magnetic and rotation energies of the star. 
Lander \& Jones (2011) also suggested that the axial-led Alfv\'en modes could be unstable.

For mixed poloidal and toroidal magnetic fields, coupled
spheroidal and toroidal velocity fields have to be considered to describe global oscillations of the magnetized stars, even for axisymmetric oscillations.
Note that 
assuming mixed poloidal and toroidal
fields for the modal analyses is favorable from a magnetic stability point of view since pure poloidal and pure toroidal magnetic field configurations are known to be unstable (e.g., Tayler 1973; Markey \& Tayler 1973, 1974; Yoshida \& Eriguchi 2006;
Yoshida, Yoshida, \& Eriguchi 2006; Akg\"un et al 2013; Herbric \& Kokkotas 2017).
Colaiuda \& Kokkotas (2012) calculated axisymmetric toroidal oscillations of
neutron stars for such a mixed field configuration.
They found that the oscillation spectra of the toroidal modes in the core
are significantly modified, losing their continuum character, by introducing a toroidal field component and
that the crustal torsional modes now become long-living oscillations. 
This finding may be similar to the finding by van Hoven \& Levin (2011, 2012) who suggested the existence of
discrete modes in the gaps between frequency continua, using a spectral method. 
Using MHD simulations, Gabler et al. (2013) discussed axisymmetric toroidal modes of magnetized stars
assuming various magnetic field configurations, but they did not
find any long-lived discrete crustal modes in the gap
between Alfv\'en continua in the core.

In this paper, instead of MHD simulations of small amplitude oscillations,
we carry out normal mode analyses 
of a magnetized neutron star which possesses a mixed poloidal and toroidal magnetic field for axisymmetric oscillations.
In normal mode analysis,
the time dependence of the oscillations is given by the factor ${\rm e}^{\rmi \omega t}$ 
and we look for the oscillation frequency $\omega$ as an eigenvalue to the set of linear differential equations that govern the oscillations. 
This paper will belong to a series of studies of normal modes of magnetized neutron stars (Lee 2007, 2008, 2010, 2018; Asai \& Lee 2014; Asai, Lee, \& Yoshida 2015, 2016).
The frequency ranges of magnetic modes obtained by normal mode calculations are similar to those
by MHD simulations.
However, the results obtained by normal mode analyses and MHD simulations are not necessarily fully consistent with each other.
For example, although Lee (2008) and Asai \& Lee (2014) obtained discrete toroidal magnetic normal modes of neutron stars for a poloidal magnetic field, MHD simulations for axisymmetric toroidal oscillations do not necessarily support the existence of such discrete magnetic normal modes.
In this paper, employing a polytrope of the index $n=1$ as the background model, we compute axisymmetric normal modes of a neutron star possessing a mixed poloidal and toroidal magnetic field.
Note that we ignore the existence of
a solid crust in neutron stars for simplicity.
\S 2 describe the method of solution we employ in this paper.
\S 3 gives numerical results and \S 4 is for conclusions.
In the Appendices A and B, we give the derivations of the equilibrium magnetic fields 
and of the outer boundary conditions.
In the Appendix C, we also briefly revisit the problem of pure toroidal magnetic modes of a neutron star with a pure poloidal field,
applying two different outer boundary conditions.

\section{Method of solution}

\subsection{Equilibrium State}

The magneto-hydrostatic equilibrium may be given by
\be
{1\over\rho}\nabla p+\nabla\Phi-{1\over 4\pi\rho}\left(\nabla\times\pmb{B}\right)\times\pmb{B}=0,
\ee
where $\rho$ is the mass density, $p$ is the pressure, $\Phi$ is the gravitational potential, and
$\pmb{B}$ denotes the magnetic field.
As discussed in the Appendix, assuming the $\phi$ component of the Lorentz force $\left(\nabla\times\pmb{B}\right)\times\pmb{B}/4\pi$ vanishes identically, 
we may derive an axisymmetric equilibrium magnetic field $\pmb{B}$ in the star given by (e.g., Colaiuda et al 2008; see also the Appendix A)
\be
B_r=2f(r)\cos\theta, \quad B_\theta=-{1\over r}{dr^2f(r)\over dr}\sin\theta, \quad B_\phi=-\zeta rf(r)\sin\theta,
\label{eq:emagf}
\ee
where $\zeta$ is a constant and the function $f$ is a solution to the differential equation given by
\be
{d^2f\over dr^2}+{4\over r}{df\over dr}+\zeta^2 f=-4\pi c_0\rho,
\label{eq:ffunc}
\ee
and $f(r)=\alpha_0+O(r^2)$ is assumed at the stellar centre, where $c_0$ and $\alpha_0$ are constants.
With the magnetic fields given by equations (\ref{eq:emagf}), the magneto-hydrostatic equation may reduce to
\be
{1\over\rho}\nabla p+\nabla\left(\Phi-c_0fr^2\sin^2\theta\right)=0,
\label{eq:mhydrostat}
\ee
where the term $c_0fr^2\sin^2\theta$ is responsible for the deviation of the equilibrium from spherical symmetry.
In this paper, we ignore the term $c_0fr^2\sin^2\theta$ so that $p$, $\rho$, and $\Phi$
depend only on the radial distance $r$ from the centre.
This approximation makes it possible to integrate equation (\ref{eq:ffunc}), given
a spherical symmetric model, which is provided by a polytrope.
See Asai et al (2016),
who took account of the equilibrium deformation caused by toroidal magnetic fields to calculate
various oscillation modes.

Assuming that $\rho=0$ and $\zeta=0$ for $r>R$ with $R$ being the radius of the star, 
we have the exterior solution $f^{\rm ex}$
given by $f^{\rm ex}=\mu_b/r^3$, where $\mu_b$ is the magnetic dipole moment of the star.
The constants $c_0$ and $\alpha_0$ are determined so that the interior solution $f$ and $df/dr$
are matched with the exterior solution $f^{\rm ex}$ and $df^{\rm ex}/dr$ at the surface $r=R$.
Although $B_r$ and $B_\theta$ are continuous at the stellar surface, $B_\phi$ is discontinuous
at $r=R$, which results in the existence of a surface current $J_\theta$ (see the Appendix {A}).

We note that for the magnetic field given by equation (\ref{eq:emagf}) the poloidal component $\pmb{B}_{\rm pl}=B_r\pmb{e}_r+B_\theta\pmb{e}_\theta$ is antisymmetric with respect to
the equator and the toroidal component $\pmb{B}_{\rm tr}=B_\phi\pmb{e}_\phi$ is symmetric, and hence that
the field $\pmb{B}=\pmb{B}_{\rm pl}+\pmb{B}_{\rm tr}$ has no definite parity with respect to the equator.

\subsection{Oscillation Equations}

Assuming the spherical symmetric structure of the star, 
neglecting the equilibrium deformation due to the magnetic field,
we linearize the ideal MHD equations to obtain
\be
\rho'+\nabla\cdot\left(\rho\pmb{\xi}\right)=0,
\label{eq:conteq}
\ee
\be
-\omega^2\pmb{\xi}
+{1\over\rho}\nabla p'-{\rho'\over\rho^2}{dp\over dr}\pmb{e}_r
-{1\over 4\pi \rho}\left[\left(\nabla\times\pmb{B}'\right)\times\pmb{B}+\left(\nabla\times\pmb{B}\right)\times
\pmb{B}'\right]=0,
\label{eq:eqmot}
\ee
\be
\pmb{B}'=\nabla\times\left(\pmb{\xi}\times\pmb{B}\right),
\label{eq:inductioneq}
\ee
\be
{\rho'\over\rho}={1\over\Gamma_1}{p'\over p}-rA{\xi_r\over r},
\label{eq:rhoprime}
\ee
where $\pmb{\xi}$ is the displacement vector and
the prime $(')$ indicates Euler perturbations, and we have assumed that the time dependence of the 
perturbations is given by the factor ${\rm e}^{\rmi \omega t}$ with $\omega$ being the oscillation frequency, and
\be
\Gamma_1=\left({\partial\ln p\over \partial\ln \rho}\right)_{\rm ad},
\quad rA={d\ln\rho\over d\ln r}-{1\over\Gamma_1}{d\ln p\over d\ln r},
\ee
and $A$ is called the Schwarzschild discriminant.
Note that we have applied the Cowling approximation neglecting $\Phi'$, the Eulerian perturbation of
the gravitational potential.

We may write the $\phi$ component of the perturbed induction equation (\ref{eq:inductioneq}) as
\be
B^\prime_\phi={1\over r}\left[{\partial\over\partial r}r\left(\xi_\phi B_r-\xi_r B_\phi\right)-{\partial\over\partial\theta}\left(\xi_\theta B_\phi-\xi_\phi B_\theta\right)\right].
\ee
Since $B_r$ is an odd function of $\cos\theta$ and $B_\theta$ and $B_\phi$ are even functions,
for $B^\prime_\phi$ to have a definite parity,
$\xi_\theta$ and $\xi_\phi$ must be antisymmetric about the equator for symmetric $\xi_r$
or symmetric for antisymmetric $\xi_r$.
Similarly, using the $\phi$ component of the equation of motion (\ref{eq:eqmot}) given by
\be
\omega^2\xi_\phi=-{1\over 4\pi\rho}\left[{1\over r}{\partial\over\partial\theta}B_\theta B^\prime_\phi
+{B_r\over r}{\partial\over\partial r}rB^\prime_\phi+\zeta\left(-B_rB^\prime_\theta+B_\theta B^\prime_r\right)\right],
\ee
we may conclude that for $\xi_\phi$ to have a definite parity,
$B^\prime_\theta$ and $B^\prime_\phi$ must be antisymmetric about the equator for symmetric $B^\prime_r$
or symmetric for antisymmetric $B^\prime_r$.

Because of the Lorentz terms in equation (\ref{eq:eqmot}), separation of variables using a single spherical harmonic function $Y_l^m(\theta,\phi)$ is not possible
to represent the perturbations of magnetized stars.
We therefore employ the series expansion in terms of spherical harmonic functions $Y_l^m$
for the perturbations.
For the displacement vector $\pmb{\xi}$ of axisymmetric modes of $m=0$, we write
\be
{\pmb{\xi}\over r}=\sum_j^{j_{\rm max}}\left[\pmb{e}_r S_{l_j}(r)+\pmb{e}_\theta H_{l_j}(r){\partial\over\partial\theta}
-\pmb{e}_\phi T_{l_j}(r){\partial\over\partial \theta}\right]Y_{l_j}^0(\theta,\phi){\rm e}^{\rmi\omega t},
\label{eq:expansionxi}
\ee
and for the perturbed magnetic fields $\pmb{B}'$
\be
{\pmb{B}^\prime\over f}=\sum_j^{j_{\rm max}}\left[\pmb{e}_r b^S_{l'_j}(r)+\pmb{e}_\theta b^H_{l'_j}(r){\partial\over\partial\theta}
-\pmb{e}_\phi b^T_{l'_j}(r){\partial\over\partial \theta}\right]Y_{l'_j}^0(\theta,\phi){\rm e}^{\rmi\omega t},
\label{eq:expansionb}
\ee
where $\pmb{e}_r$, $\pmb{e}_\theta$, and $\pmb{e}_\phi$ are the orthonormal vectors in the $r$, $\theta$, and $\phi$ directions, respectively, 
and for the pressure perturbation, $p'$, we write
\be
p'=\sum_j^{j_{\rm max}}p'_{l_j}(r)Y_{l_j}^0(\theta,\phi){\rm e}^{\rmi\omega t}.
\label{eq:expansionp}
\ee
Note that the modes represented by the series expansions
are separated into two groups according to $(l_j,l'_j)$, that is,
$(l_j,l'_j)=(2j-2,2j-1)$ for one group and $(l_j,l'_j)=(2j-1,2j-2)$ for the other where $j=1,~ 2,~3,\cdots$.
In this paper we call the former even modes and the latter odd modes since the pressure perturbation $p'$
is symmetric with respect to the equator for the former and antisymmetric for the latter.

If we define the spheroidal $\pmb{\xi}_{\rm SH}$ and toroidal $\pmb{\xi}_{\rm T}$ components of $\pmb{\xi}$ as
\be
{\pmb{\xi}_{\rm SH}\over r}=\sum_j^{j_{\rm max}}\left(\pmb{e}_r S_{l_j}+\pmb{e}_\theta H_{l_j}{\partial\over\partial\theta}
\right)Y_{l_j}^0{\rm e}^{\rmi\omega t}, \quad 
{\pmb{\xi}_{\rm T}\over r}=-\pmb{e}_\phi\sum_j^{j_{\rm max}}
 T_{l_j}{\partial\over\partial \theta}Y_{l_j}^0{\rm e}^{\rmi\omega t},
 \label{eq:exxi}
\ee
and the spheroidal $\pmb{B}^\prime_{\rm SH}$ and toroidal $\pmb{B}^\prime_{\rm T}$ components of $\pmb{B}^\prime$ as
\be
{\pmb{B}^\prime_{\rm SH}\over f}=\sum_j^{j_{\rm max}}\left(\pmb{e}_r b^S_{l'_j}+\pmb{e}_\theta b^H_{l'_j}{\partial\over\partial\theta}
\right)Y_{l_j}^0{\rm e}^{\rmi\omega t}, \quad 
{\pmb{B}^\prime_{\rm T}\over f}=-\pmb{e}_\phi\sum_j^{j_{\rm max}}
 b^T_{l'_j}{\partial\over\partial \theta}Y_{l'_j}^0{\rm e}^{\rmi\omega t},
 \label{eq:exb}
\ee
$\pmb{\xi}_{\rm SH}$ and $\pmb{B}^\prime_{\rm T}$ are symmetric about the equator
and $\pmb{\xi}_{\rm T}$ and $\pmb{B}^\prime_{\rm SH}$ are antisymmetric for $(l_j,l'_j)=(2j-2,2j-1)$
and vice versa for $(l_j,l'_j)=(2j-1,2j-2)$.
When $\zeta=0$, the spheroidal components $(\pmb{\xi}_{\rm SH},\pmb{B}^\prime_{\rm SH})$ are decoupled from the toroidal
ones $(\pmb{\xi}_{\rm T},\pmb{B}^\prime_{\rm T})$, and 
we call
$(\pmb{\xi}_{\rm SH},\pmb{B}^\prime_{\rm SH})$ even spheroidal modes for $(l_j,l'_j)=(2j-2,2j-1)$
and odd spheroidal modes for $(l_j,l'_j)=(2j-1,2j-2)$ according to the parity of $\pmb{\xi}_{\rm SH}$, 
while we call $(\pmb{\xi}_{\rm T},\pmb{B}^\prime_{\rm T})$
odd toroidal modes for $(l_j,l'_j)=(2j-2,2j-1)$ and even toroidal modes for $(l_j,l'_j)=(2j-1,2j-2)$, 
according to the parity of $\pmb{\xi}_{\rm T}$.
In this paper, we use this terminology even for $\zeta\not=0$.
It is important to note that for the perturbations as defined by equations
(\ref{eq:expansionxi}), (\ref{eq:expansionb}), and (\ref{eq:expansionp}), 
we consider that even spheroidal modes $(\pmb{\xi}_{\rm SH},\pmb{B}^\prime_{\rm SH})$ are coupled with odd toroidal modes $(\pmb{\xi}_{\rm T},\pmb{B}^\prime_{\rm T})$ and
vice versa when $\zeta\not=0$.
We thus note that the vectors $\pmb{\xi}$ and $\pmb{B}'$ as defined by the expansions (\ref{eq:expansionxi}) and (\ref{eq:expansionb}) do not have a definite parity about the equator.

Substituting the series expansions into the perturbed basic equations (\ref{eq:conteq}) to (\ref{eq:rhoprime}),
we obtain a set of linear ordinary differential equations for the expansion coefficients.
For the dependent variables defined as
\be
\pmb{y}_1=\left({S_{l_j}}\right), \quad \pmb{y}_2=\left({p'_{l_j}\over \rho g r}\right), \quad
\pmb{y}_3=\left(H_{l_j}\right), \quad \pmb{y}_4=\left(b^H_{l'_j}\right),
\quad \pmb{y}_5=\left(T_{l_j}\right), \quad \pmb{y}_6=\left(b^T_{l'_j}\right),\quad \pmb{b}^S=\left(b^S_{l'_j}\right),
\ee
we obtain 
\be
r{{\rm d}\pmb{y}_1\over {\rm d}r}=\left({V\over \Gamma_1}-3\right)\pmb{y}_1-{V\over\Gamma_1}\pmb{y}_2+\pmbmt{\Lambda}_0\pmb{y}_3,
\label{eq:y1}
\ee
\begin{eqnarray}
r{{\rm d}\pmb{y}_2\over {\rm d}r}
&=&\left[\left(c_1\bar\omega^2+rA\right)\pmbmt{1}-{1\over 2}\left({{\cal Q}\over f}\right)^2
{\cal S}\pmbmt{C}_0\pmbmt{B}_1^{-1}\pmbmt{W}_0\pmbmt{W}_1
\right]\pmb{y}_1
-\left[\left(rA+U-1\right)\pmbmt{1}+{1\over 2}{{\cal Q}\over f}\pmbmt{C}_0\pmbmt{B}_1^{-1}\pmbmt{\Lambda}_0\right]\pmb{y}_2\nonumber\\
&&+{{\cal Q}\over f}\left[{1\over 2}c_1\bar\omega^2\pmbmt{C}_0\pmbmt{B}_1^{-1}\pmbmt{\Lambda}_0
-{\cal S}\pmbmt{C}_0\pmbmt{B}_1^{-1}\pmbmt{W}_0\pmbmt{B}_0
\right]\pmb{y}_3
-{\cal S}\pmbmt{C}_0\pmb{y}_4
+{1\over 2}\zeta_0xc_1\bar\omega^2\pmbmt{C}_0\pmbmt{B}_1^{-1}\pmbmt{\Lambda}_0\pmb{y}_5,
\label{eq:y2}
\end{eqnarray}
\begin{eqnarray}
r{{\rm d}\pmb{y}_3\over {\rm d}r}={1\over 2}{{\cal Q}\over f}\left(3-{V\over\Gamma_1}+{{\cal Q}\over f}-{d\ln {\cal Q}/f\over d\ln r}\right)
\pmbmt{B}_0^{-1}\pmbmt{W}_1\pmb{y}_1+{1\over 2}{V\over\Gamma_1}{{\cal Q}\over f}\pmbmt{B}_0^{-1}\pmbmt{W}_1\pmb{y}_2
+{{\cal Q}\over f}\left(\pmbmt{1}-{1\over 2}\pmbmt{B}_0^{-1}\pmbmt{W}_1\pmbmt{\Lambda}_0\right)\pmb{y}_3
+{1\over 2}\pmbmt{B}_0^{-1}\pmbmt{\Lambda}_1\pmb{y}_4,
\label{eq:y3}
\end{eqnarray}
\begin{eqnarray}
r{{\rm d}\pmb{y}_4\over {\rm d}r}
&=&{{\cal Q}\over f}\left(\pmbmt{1}+{1\over 2}{{\cal R}\over f}\pmbmt{B}_1^{-1}\pmbmt{W}_0\right)\pmbmt{W}_1\pmb{y}_1
+{1\over 2\beta}\pmbmt{B}_1^{-1}\pmbmt{\Lambda}_0\pmb{y}_2
+\left[2\left(\pmbmt{1}+{1\over 2}{{\cal R}\over f}\pmbmt{B}_1^{-1}\pmbmt{W}_0\right)\pmbmt{B}_0-{1\over 2\beta}c_1\bar\omega^2\pmbmt{B}_1^{-1}\pmbmt{\Lambda}_0\right]\pmb{y}_3-{d\ln rf\over d\ln r}\pmb{y}_4\nonumber\\
&&+{1\over 2}\zeta_0x\left(\pmbmt{B}_1^{-1}\pmbmt{W}_0\pmbmt{\Lambda}_1+2\pmbmt{1}\right)\pmb{y}_6,
\label{eq:y4}
\end{eqnarray}
\begin{eqnarray}
r{{\rm d}\pmb{y}_5\over {\rm d}r}
=-{{\cal Q}\over f}\left({1\over 2}\pmbmt{B}_0^{-1}\pmbmt{W}_1\pmbmt{\Lambda}_0-\pmbmt{1}\right)\pmb{y}_5
+{1\over 2}\pmbmt{B}_0^{-1}\pmbmt{\Lambda}_1\pmb{y}_6-{1\over 2}\zeta_0x\pmbmt{B}_0^{-1}\pmbmt{W}_1\left[\left({V\over\Gamma_1}-2-{Q\over f}\right)\pmb{y}_1-{V\over\Gamma_1}\pmb{y}_2\right],
\label{eq:y5}
\end{eqnarray}
\begin{eqnarray}
r{{\rm d} \pmb{y}_6\over {\rm d}r}
=-{1\over 2\beta}c_1\bar\omega^2\pmbmt{B}_1^{-1}\pmbmt{\Lambda}_0\pmb{y}_5
-\left({1\over 2}{{\cal Q}\over f}\pmbmt{B}_1^{-1}\pmbmt{W}_0\pmbmt{\Lambda}_1+{d\ln fr\over d\ln r}\pmbmt{1}\right)\pmb{y}_6
-{\zeta_0\over 2}x\left[\left({Q\over f}\right)^2\pmbmt{B}_1^{-1}\pmbmt{W}_0\pmbmt{W}_1\pmb{y}_1+2{Q\over f}\pmbmt{B}_1^{-1}\pmbmt{W}_0\pmbmt{B}_0\pmb{y}_3+2\pmb{y}_4\right],
\label{eq:y6}
\end{eqnarray}
where
\be
\bar\omega={\omega\over\sigma_0}, \quad \sigma_0=\sqrt{GM\over R^3}, \quad U={d\ln M_r\over d\ln r}, \quad 
V=-{d\ln p\over d\ln r},
\ee
\be
c_1={(r/R)^3\over M_r/M}, \quad g={GM_r\over r^2}, \quad M_r=\int_0^r4\pi r^2\rho dr, 
\ee
\be
{\cal Q}=-{1\over r}{d(r^2f)\over dr},
\quad {\cal R}
=-r^2\left({d^2f\over dr^2}+{4\over r}{df\over dr}\right), 
\quad {\cal S}=\beta\left({{\cal R}\over f}-\zeta_0^2x^2\right), 
\ee
\be
\beta={f^2\over 4\pi pV}, \quad \zeta_0=\zeta R, \quad x={r\over R},
\ee
and $\pmbmt{1}$ denotes the identity matrix, and 
the matrices $\pmbmt{B}_0$, $\pmbmt{B}_1$, $\pmbmt{W}_0$, and $\pmbmt{W}_1$ are defined as
\be
\pmbmt{B}_0=\pmbmt{Q}_1\pmbmt{\Lambda}_0+\pmbmt{C}_1, \quad \pmbmt{B}_1=\pmbmt{Q}_0\pmbmt{\Lambda}_1+\pmbmt{C}_0, \quad \pmbmt{W}_0=2\pmbmt{Q}_0+\pmbmt{C}_0, \quad \pmbmt{W}_1=2\pmbmt{Q}_1+\pmbmt{C}_1,
\ee
and the definition of the matrices $\pmbmt{Q}_0$, $\pmbmt{Q}_1$, $\pmbmt{C}_0$, $\pmbmt{C}_1$, $\pmbmt{\Lambda}_0$, and $\pmbmt{\Lambda}_1$ is given in Yoshida \& Lee (2000).
We note that
\be
\pmbmt{W}_1\pmbmt{\Lambda}_0=\pmbmt{\Lambda}_1\pmbmt{C}_1, \quad \pmbmt{W}_0\pmbmt{\Lambda}_1=\pmbmt{\Lambda}_0\pmbmt{C}_0, \quad \pmbmt{W}_0=-\pmbmt{C}_1^T, \quad \pmbmt{W}_1=-\pmbmt{C}_0^T,
\ee
and that although the quantity $\beta$ diverges both at the centre and at the surface, the quantity $\cal S$ remains finite
in the interior.
It is important to note that since the expansion coefficient $H_{l_1}=H_0$, for example, vanishes identically for even modes, 
we take $H_{l_2}$ to $H_{l_{j_{{\rm max}}+1}}$ as dependent variables and hence we have to redefine the matrices given above accordingly (see, e.g., Lee 2008).
To simplify the set of differential equations, we have used
\be
\pmb{b}^S={Q\over f}\pmbmt{W}_1\pmb{y}_1+2\pmbmt{B}_0\pmb{y}_3,
\label{eq:induction_r}
\ee
which comes from the radial component of the induction equation (\ref{eq:inductioneq}), and
\be
r{{\rm d}\pmb{b}^S\over {\rm d}r}={Q\over f}\pmb{b}^S+\pmbmt{\Lambda}_1\pmb{y}_4,
\label{eq:nablab}
\ee
which comes from $\nabla\cdot\pmb{B}'=0$.
Note that the parameter $\zeta_0$ represents the ratio of the toroidal component of the magnetic field
to the poloidal one and that the parameter $\zeta_0\not=0$ couples the spheroidal modes and toroidal modes, that is, when $\zeta_0=0$
the set of differential equations from (\ref{eq:y1}) to (\ref{eq:y6}) are decoupled into those for spheroidal modes
described by the variables $(\pmb{y}_1, \pmb{y}_2, \pmb{y}_3, \pmb{y}_4)$ and for toroidal modes by
$(\pmb{y}_5, \pmb{y}_6)$ (see, e.g., Lee 2007).

Imposing appropriate boundary conditions at the centre and at the
surface of the star, we solve equations $(\ref{eq:y1})$ to $(\ref{eq:y6})$ as an eigenvalue problem for the oscillation frequency $\omega$.
The inner boundary conditions are regularity conditions for the variables $\pmb{\xi}$, $p'$, and $\pmb{B}'$
at the stellar centre.
The surface boundary conditions are given by
\begin{eqnarray}
-\pmb{y}_1+\pmb{y}_2+\zeta_0\beta\left({{\cal Q}\over f}+2\right)\pmbmt{Q}_0\pmbmt{C}_1\pmb{y}_5=0,
\end{eqnarray}
\be
\left[{{\cal Q}\over f}\pmbmt{W}_1+\zeta_0^2\pmbmt{L}^+\pmbmt{C}_0^{-1}\left(\pmbmt{1}-\pmbmt{Q}_0\pmbmt{Q}_1\right)\right]\pmb{y}_1+2\pmbmt{B}_0\pmb{y}_3+\pmbmt{L}^+\pmb{y}_4
+\zeta_0\left(\pmbmt{C}_1-\pmbmt{L}^+\pmbmt{C}_0^{-1}\pmbmt{Q}_0\pmbmt{C}_1\right)\pmb{y}_5=0,
\ee
and
\be
-\zeta_0\pmbmt{C}_0^{-1}\left[\left({{\cal Q}\over f}+1\right)\left(\pmbmt{1}-\pmbmt{Q}_0\pmbmt{Q}_1\right)\pmb{y}_1-\pmbmt{Q}_0\pmbmt{C}_1\pmb{y}_3\right]+\pmb{y}_6=0,
\ee
where $(\pmbmt{L}^+)_{ij}=(l'_i+1)\delta_{ij}$ and the derivation of the surface boundary conditions are given in the Appendix {B}.

\subsection{Energy Equation}

It is useful to derive an energy equation for the perturbations associated with adiabatic flows.
We start with introducing the force operator $\pmb{F}(\pmb{\xi})$ defined by
\be
\pmb{F}(\pmb{\xi})=-\rho\nabla\Phi'-\nabla p'+{\rho'\over\rho}\nabla p+{1\over 4\pi}\left[\left(\nabla\times\pmb{B}^\prime\right)\times\pmb{B}+\left(\nabla\times\pmb{B}\right)\times\pmb{B}^\prime\right],
\ee
which satisfies the perturbed equation of motion 
\be
\rho\partial^2\pmb{\xi}/\partial t^2=\pmb{F}(\pmb{\xi}).
\label{eq:perteom}
\ee
The energy equation may be given by multiplying the perturbed equation of motion (\ref{eq:perteom})
by $\pmb{\xi}^*$,  
the complex conjugate of the displacement vector $\pmb{\xi}$.
Assuming the time dependency of the perturbations is given by ${\rm e}^{\rmi\omega t}$,
we obtain
\begin{eqnarray}
\omega^2\pmb{\xi}^*\cdot\pmb{\xi}&=&
{1\over\Gamma_1}{p'\over\rho}{p'^*\over p}+{p\over\rho}\left(\pmb{\xi}^*\cdot\nabla\ln p\right)\left(\pmb{\xi}\cdot
\pmb{A}\right)-{\nabla\Phi'\cdot\nabla\Phi'^*\over 4\pi G\rho}
-{1\over 4\pi\rho}\pmb{\xi}^*\cdot\left[\left(\nabla\times\pmb{B}'\right)\times\pmb{B}+\left(\nabla\times\pmb{B}\right)\times\pmb{B}'\right]
\nonumber\\
&&
+{1\over\rho}\nabla\cdot\left[\pmb{\xi}^*\left(p'+\rho\Phi'\right)
+{\Phi'\nabla\Phi'^*\over 4\pi G}\right]
\end{eqnarray}
where
\be
\pmb{A}=\nabla\ln\rho-{1\over\Gamma_1}\nabla\ln p.
\ee
When we assume spherical symmetry of the equilibrium configuration, neglecting $\pmb{J}\times\pmb{B}$ term
in the magneto-hydrostatic equilibrium,
we may obtain, using the relation $\pmb{B}'=\nabla\times\left(\pmb{\xi}\times\pmb{B}\right)$,
\begin{eqnarray}
\omega^2\pmb{\xi}^*\cdot\pmb{\xi}={1\over\rho}\nabla\cdot\left(\pmb{\xi}^* p'\right)+{V\over\Gamma_1}{p^{\prime *}
\over\rho gr}{p\over \rho}-grA{\xi_r^*\xi_r\over r}+{1\over 4\pi\rho}\left[\left|\pmb{B}'\right|^2
+\left(\nabla\times\pmb{B}\right)\cdot\left(\pmb{\xi}^*\times\pmb{B}'\right)\right]
-{1\over 4\pi\rho}\nabla\cdot\left[\left(\pmb{\xi}^*\times\pmb{B}\right)\times\pmb{B}'\right],
\end{eqnarray}
where $A$ is the radial component of $\pmb{A}$ and we have applied the Cowling approximation, neglecting $\Phi'$.
Integrating over the stellar mass, we obtain
\begin{eqnarray}
\omega^2\int dV\rho\pmb{\xi}^*\cdot\pmb{\xi}&=&\int dV\rho\left\{{V\over\Gamma_1}{p'\over\rho}{p'^*\over\rho gr}
-grA{\xi_r^*\xi_r\over r}+{1\over 4\pi \rho}\left[\left|\pmb{B}'\right|^2+\left(\nabla\times\pmb{B}\right)\cdot\left(\pmb{\xi}^*\times\pmb{B}'\right)\right]\right\}\nonumber\\
&&+\int dS\pmb{n}\cdot\left[\pmb{\xi}^*p'-{1\over 4\pi}\left(\pmb{\xi}^*\times\pmb{B}\right)\times\pmb{B}'\right],
\end{eqnarray}
where $\pmb{n}$ is the unit normal vector at the surface.
Note that the surface term may be rewritten as
\be
\int dS\pmb{n}\cdot\left[\pmb{\xi}^*p'-{1\over 4\pi}\left(\pmb{\xi}^*\times\pmb{B}\right)\times\pmb{B}'\right]
=\int dS\left\{\xi_r^*\left[p'+{1\over 4\pi}\left(\pmb{B}\cdot\pmb{B}'\right)\right]-{1\over 4\pi}\left(\pmb{\xi}^*\cdot
\pmb{B}'\right)B_r\right\}.
\ee
The square of the oscillation frequency is now given by using the eigenfunctions as
\be
\omega^2={\displaystyle \int dV\rho\left\{{V\over\Gamma_1}{p'\over\rho}{p'^*\over\rho gr}
-grA{\xi_r^*\xi_r\over r}+{1\over 4\pi \rho}\left[\left|\pmb{B}'\right|^2+\left(\nabla\times\pmb{B}\right)\cdot\left(\pmb{\xi}^*\times\pmb{B}'\right)\right]\right\}+\int dS\pmb{n}\cdot\left[\pmb{\xi}^*p'-{1\over 4\pi}\left(\pmb{\xi}^*\times\pmb{B}\right)\times\pmb{B}'\right]\over\displaystyle\int dV\rho\pmb{\xi}^*\cdot\pmb{\xi}},
\label{eq:omegaint}
\ee
which we may use to cheque numerical consistency of our eigen-mode computations.
We let $\omega^2_{\rm int}$ denote the right-hand-side of equation (\ref{eq:omegaint}) and define $\delta_{\rm cs}$, using an eigenfrequency $\omega$ and $\omega_{\rm int}$ computed with the corresponding eigenfunctions, as
\be
\delta_{\rm cs}\equiv2\left|{\omega-\omega_{\rm int}\over \omega+\omega_{\rm int}}\right|.
\ee
We may consider that $\delta_{\rm cs}\ll 1$ suggests a good consistency in the computation.

\section{Numerical Results}

\begin{figure}
\begin{center}
\resizebox{0.4\columnwidth}{!}{
\includegraphics{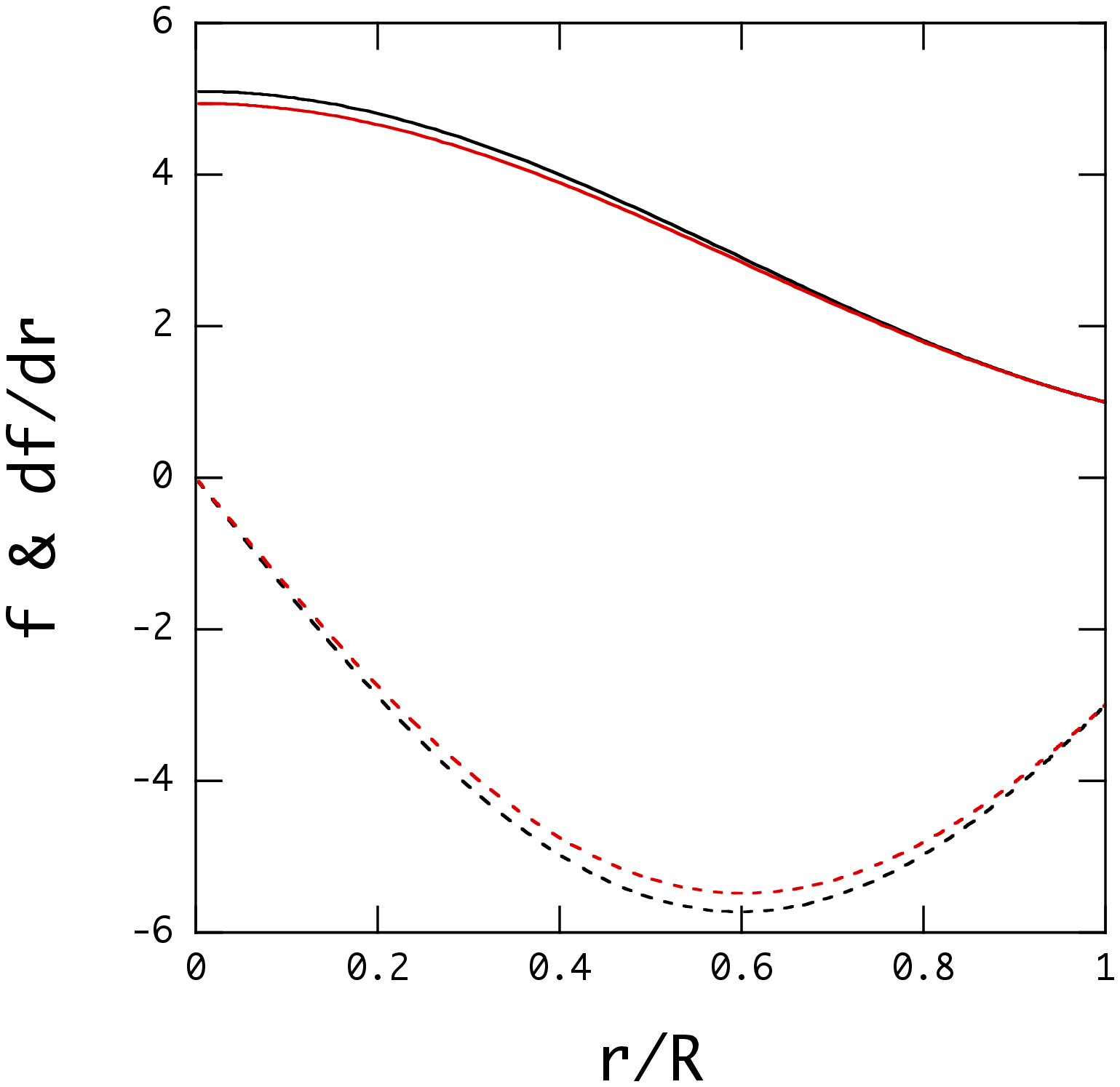}}
\end{center}
\caption{Functions $f$ (solid lines) and $df/dr$ (dotted lines) versus $x=r/R$ for a polytrope of the index $n=1$ for $\zeta_0=0.01$ (black lines) and $\zeta_0=1$ (red lines), where $f$ and $df/dr$ are normalized by $B_S\equiv \mu_b/R^3$ and $B_S/R$, respectively.
}
\label{fig:fdfdr}
\end{figure}

Ignoring the $\pmb{J}\times\pmb{B}$ term in the magneto-hydrostatic equilibrium,
we use a polytrope of the index $n=1$, which is spherical symmetric, as the background model for our modal analyses.
Here, we assume the mass $M=1.4M_\odot$ and radius $R=10^6$cm for the polytrope.
We also assume that the polytrope is isentropic, that is, $A=0$ in the entire interior of the star.

For given $\zeta_0\equiv \zeta R$, $B_S\equiv \mu_b/R^3$, and $j_{\rm max}$,
we integrate equation (\ref{eq:ffunc}) to compute the function $f$ using $\rho(r)$ from 
the polytrope, and solve the set of differential equations from (\ref{eq:y1}) to (\ref{eq:y6}) together with
the boundary conditions to obtain an eigen-solution, that is, an eigenfrequency $\omega$ and the corresponding
eigenfunctions $\pmb{y}_j$.
We usually find numerous eigen-solutions to the set of differential equations for a given $j_{\rm max}$ and
we pick up only eigen-solutions whose eigenfrequency $\omega$ well converges as $j_{\rm max}$ increases.
Good convergence is usually expected for solutions that have only a few nodes of the eigenfunctions of dominating amplitudes.
In Fig. 1, we plot the functions $f$ and $df/dr$ for $\zeta_0=0.01$ and $\zeta_0=1$.
Although the magnitudes of the parameter $\zeta_0$ are different by a factor 100, 
the difference of the functions $f$ and $df/dr$ between the two cases is not necessarily
significant.
This may be because the constant $\zeta^2\sim 0.1$ in equation (\ref{eq:ffunc}) for $\zeta_0=1$ and because the function $f$ is well determined by the outer boundary conditions that do not depend on the parameter $\zeta_0$.
Note that at $\zeta_0=0$, a magnetic field line $\pmb{B}$ given by equation (\ref{eq:emagf})
stays in a plane perpendicular to the equatorial plane, and the field lines in such a plane look like those given in Lee (2018), for example.
For $\zeta_0\not=0$, however, no field lines can stay in a plane because of $B_\phi\not=0$.
The open field lines at $\zeta_0=0$ will be twisted for $\zeta_0\not=0$ and the closed field lines
are not closed in a plane any more.

\subsection{Spheroidal Modes}

In Fig. 2 we plot $\bar\omega$ of the first two spheroidal magnetic modes of even parity (left panel) and of odd parity (right panel) as a function of $\zeta_0$ for $B_S=10^{14}$G (red dots) and $B_S=10^{15}$G (short vertical lines), 
where we have used $j_{\rm max}=14$ and $10\times\bar\omega$ is plotted for $B_S=10^{14}$G.
The figure suggests that the frequency of the spheroidal magnetic modes is almost exactly
proportional to the field strength $B_S$ for $\zeta_0\ltsim 1$.
We find $\delta_{\rm cs}\sim 10^{-3}$ for even parity modes and $\delta_{\rm cs}\sim 10^{-2}$ for
odd parity modes, that is, odd parity modes show rather poor consistency compared to even parity modes.
The figure also shows that as the parameter $\zeta_0$ increases the frequency $\bar\omega$ decreases and that
the magnetic modes sometimes suffer avoided crossings with other magnetic modes.
We find that the consistency $\delta_{\rm cs}$ becomes poor at such avoided crossings.

\begin{figure}
\begin{center}
\resizebox{0.4\columnwidth}{!}{
\includegraphics{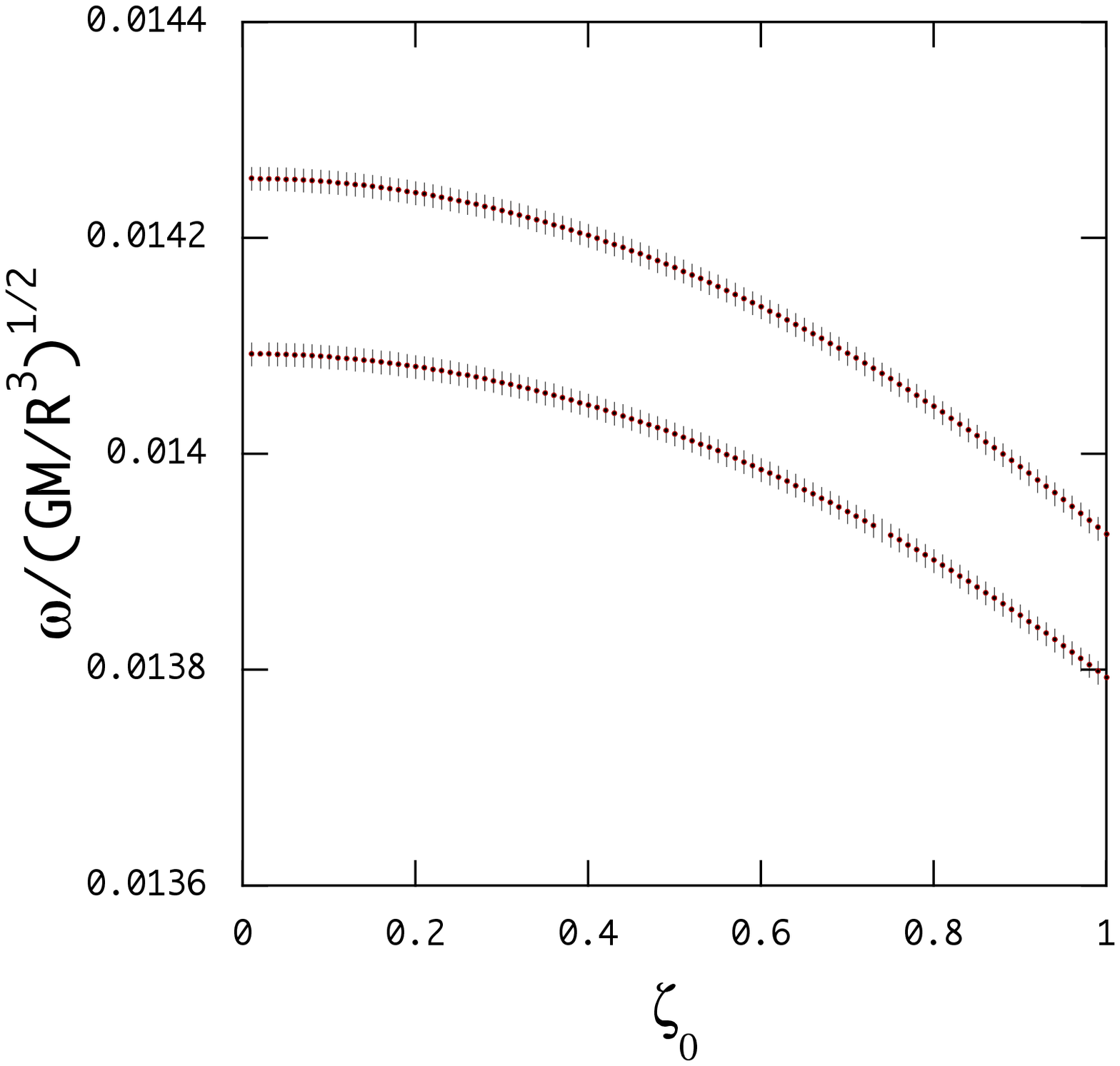}}
\resizebox{0.4\columnwidth}{!}{
\includegraphics{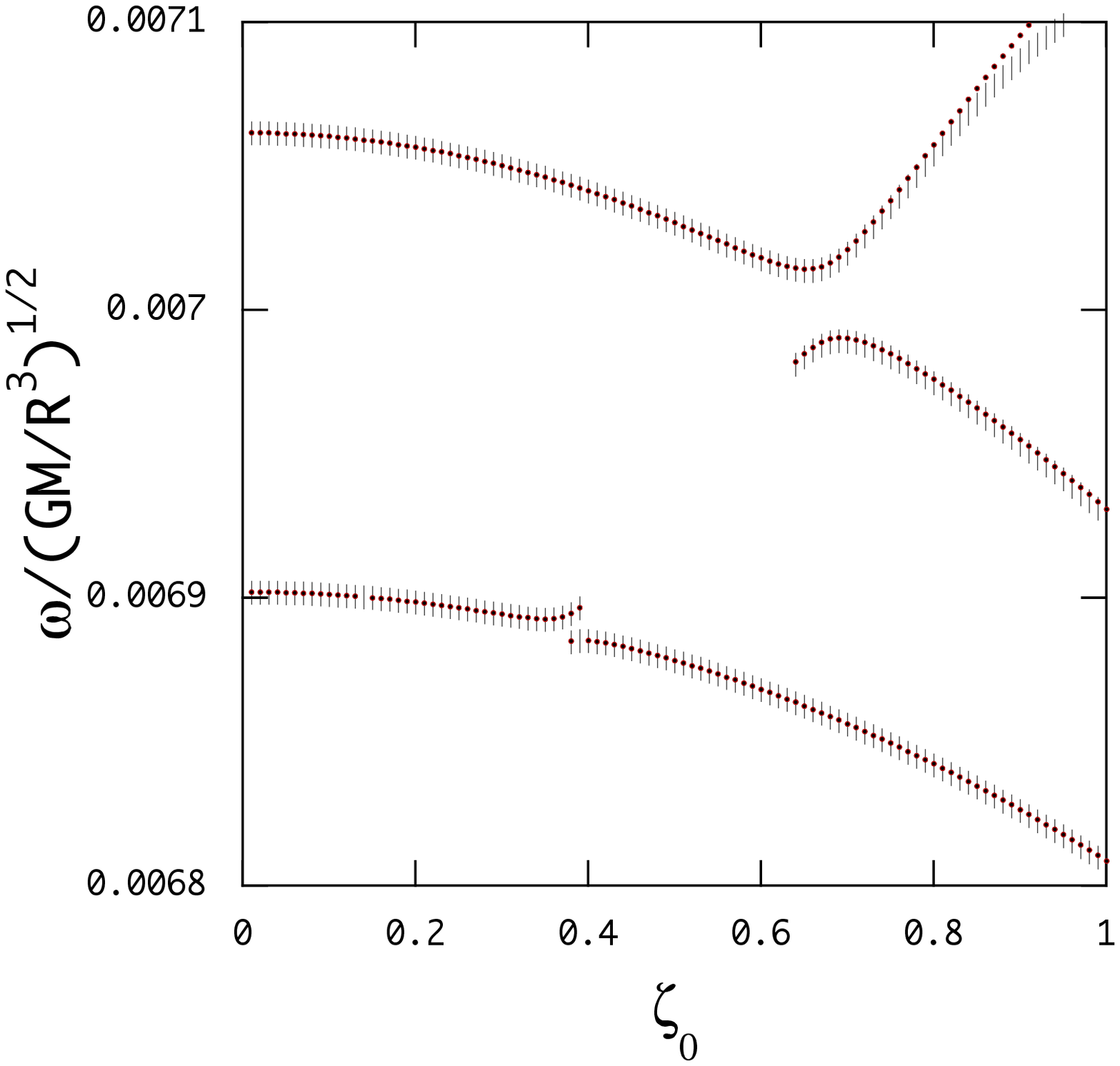}}

\end{center}
\caption{Eigenfrequency $\bar\omega$ of the first two magnetic modes of even parity (left panel) and odd parity (right panel)
versus $\zeta_0$ for $B_S=10^{15}$G (short vertical lines) and $B_S=10^{14}$G (red dots),
where the frequency $10\times \bar\omega$ is plotted for $B_S=10^{14}$G.
}
\label{fig:fdfdr}
\end{figure}

Defining the function $\hat\xi_j$ for $j=r,~\theta,~\phi$ as
\be
\hat\xi_j(r,\theta)=\xi_j(r,\theta,t){\rm e}^{-\rmi\omega t},
\ee
we plot the patterns of $\hat\xi_j(r,\theta)$ for the magnetic modes on the $x$-$z$ plane where
$x=r\sin\theta$ and $z=r\cos\theta$.
Fig.3 shows the patterns $\hat\xi_j$ of the spheroidal magnetic mode of even parity for $B_S=10^{15}$G at $\zeta_0=0.01$, where the frequency is $\bar\omega=1.426\times10^{-2}$.
For even modes, the pattern $\hat\xi_r$ is symmetric and those of $\hat\xi_\theta$ and $\hat\xi_\phi$ are
antisymmetric about the equator, that is, with respect to the $z=0$ plane.
Note also that $\hat\xi_r$ has finite amplitudes on the magnetic ($z$-) axis but
$\hat\xi_\theta$ and $\hat\xi_\phi$ have vanishing amplitudes.
The patterns $\hat\xi_r$ and $\hat\xi_\theta$ look quite similar to those computed by assuming $\zeta_0=0$
(see Lee 2018), and the amplitudes are mostly confined to the region along the magnetic axis.
$\hat\xi_\theta$ also shows wavy patterns in the region of closed magnetic fields.
The amplitude of $\xi_\phi$ (and $B'_\phi$) is proportional to $\zeta_0$ and is much smaller than $\xi_r$ and $\xi_\theta$ for $\zeta_0=0.01$, and the patterns $\hat\xi_\phi$ show short wavy features both in the region
of closed magnetic fields and in the region along the magnetic axis.
These short wavy features may suggest that the eigenfunction $\xi_\phi$ is not necessarily well
converged for $j_{\rm max}=14$.

\begin{figure}
\begin{center}
\resizebox{0.33\columnwidth}{!}{
\includegraphics{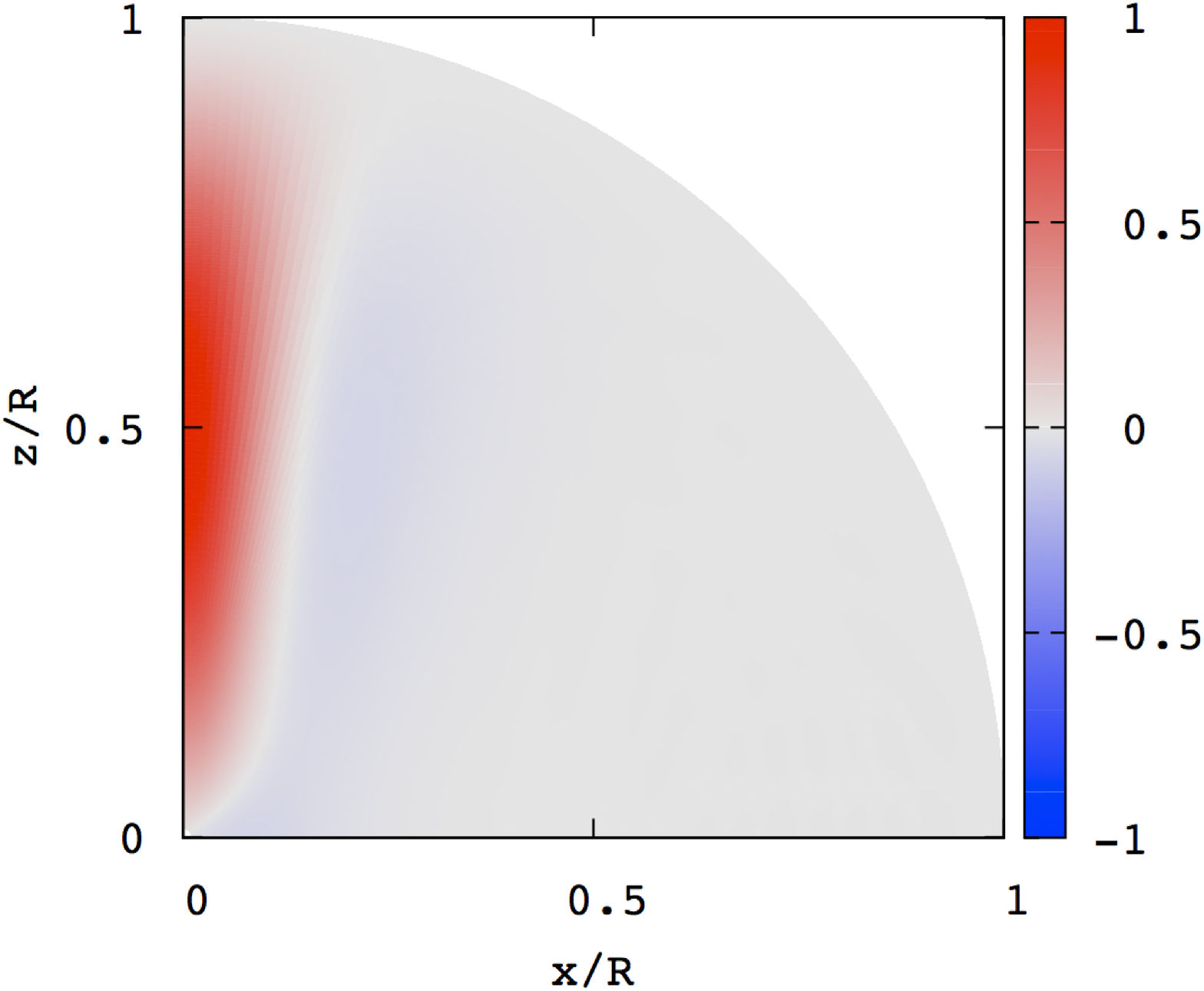}}
\resizebox{0.33\columnwidth}{!}{
\includegraphics{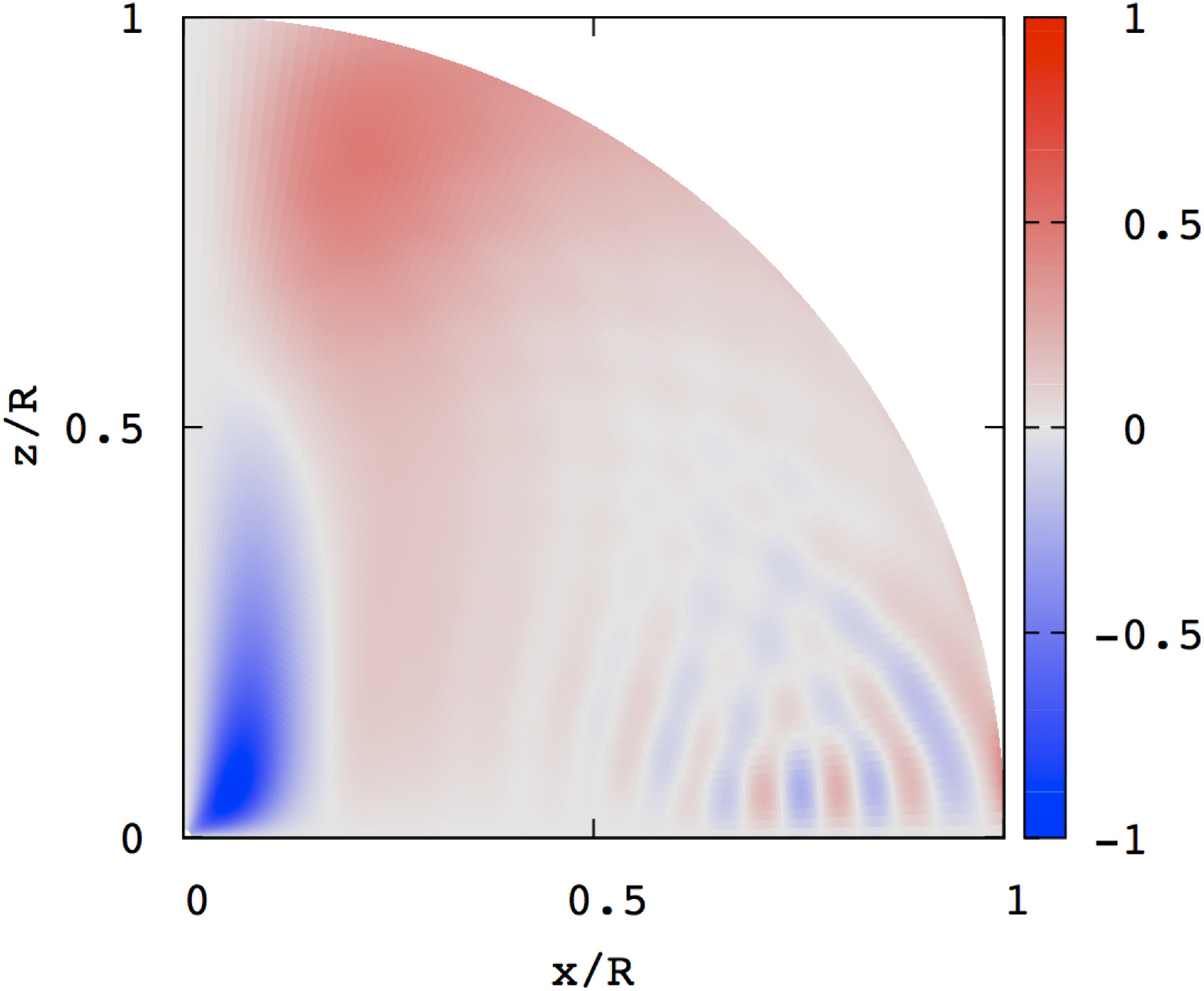}}
\resizebox{0.33\columnwidth}{!}{
\includegraphics{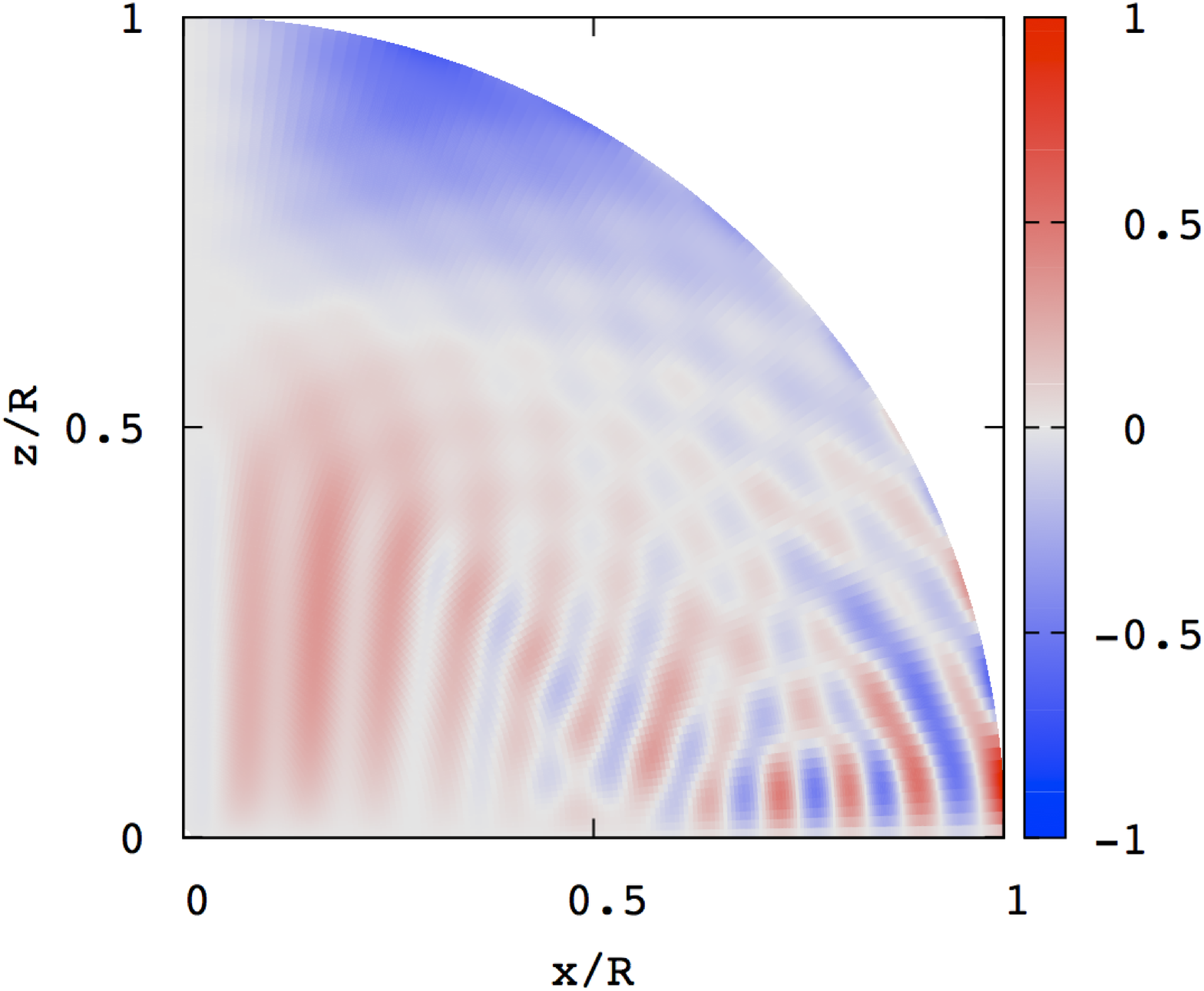}}
\end{center}
\caption{Wave patterns $\hat\xi_r$ (left panel), $\hat\xi_\theta$ (middle panel), and $\hat\xi_\phi$ (right panel) for the spheroidal magnetic mode of even parity for
$B_S=10^{15}$G at $\zeta_0=0.01$ where $\bar\omega=1.426\times10^{-2}$.
}
\label{fig:expcoeffz001}
\end{figure}

Fig.4 shows the patterns $\hat\xi_j$ for the same mode as in Fig. \ref{fig:expcoeffz001} but for $\zeta_0=1$, 
where the frequency is $\bar\omega=1.392\times10^{-2}$.
The patterns of $\hat\xi_r$ and $\hat\xi_\theta$ at $\zeta_0=1$ are almost the same as those at $\zeta_0=0.01$.
The pattern $\hat\xi_\phi$ at $\zeta_0=1$, however, is significantly different from that at $\zeta=0.01$, and
the short wavy patterns found for $\zeta_0=0.01$ disappear for $\zeta_0=1$.
Note that the amplitude of $\xi_\phi$ at $\zeta_0=1$ is comparable to those of $\xi_r$ and $\xi_\theta$.

\begin{figure}
\begin{center}
\resizebox{0.33\columnwidth}{!}{
\includegraphics{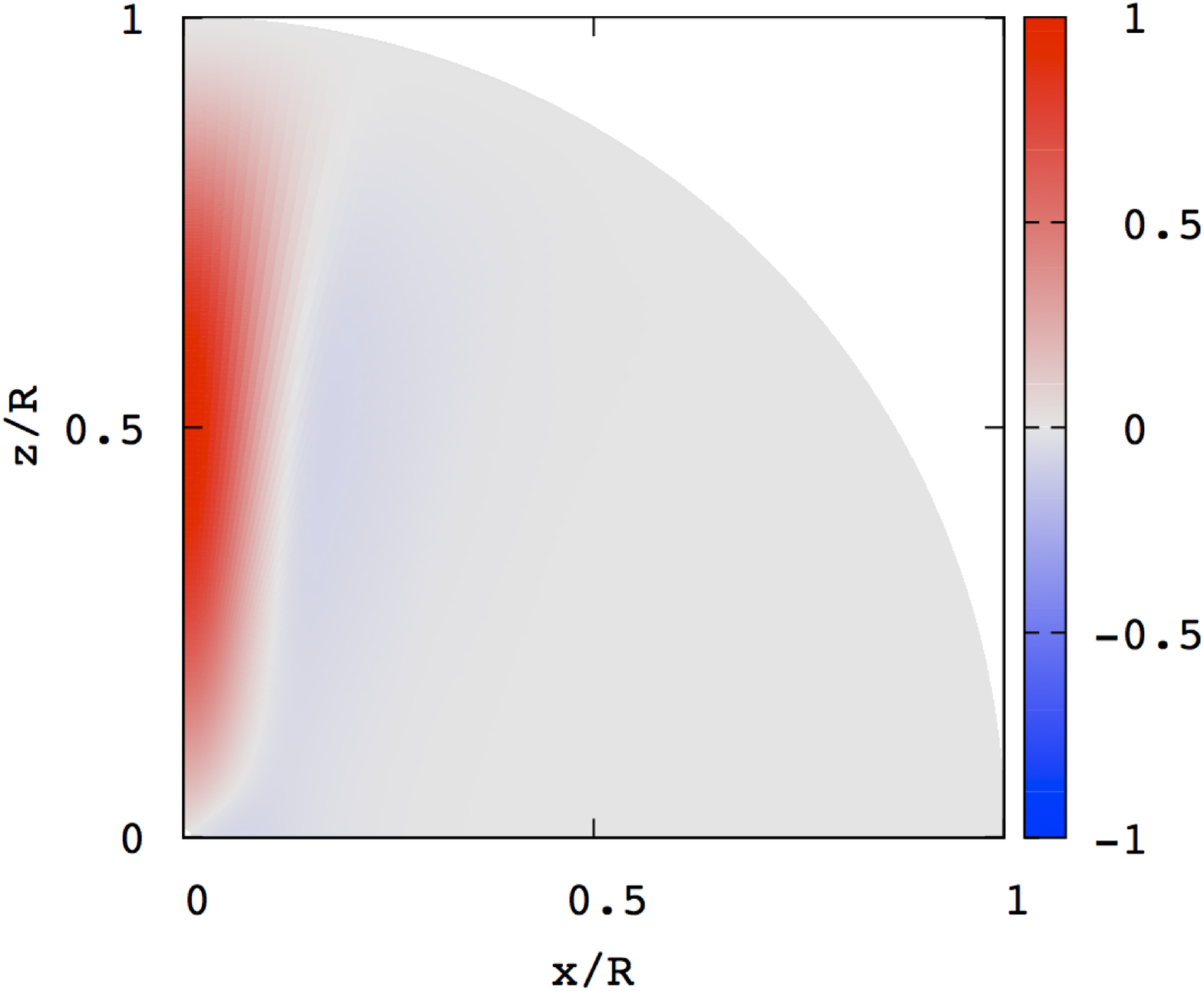}}
\resizebox{0.33\columnwidth}{!}{
\includegraphics{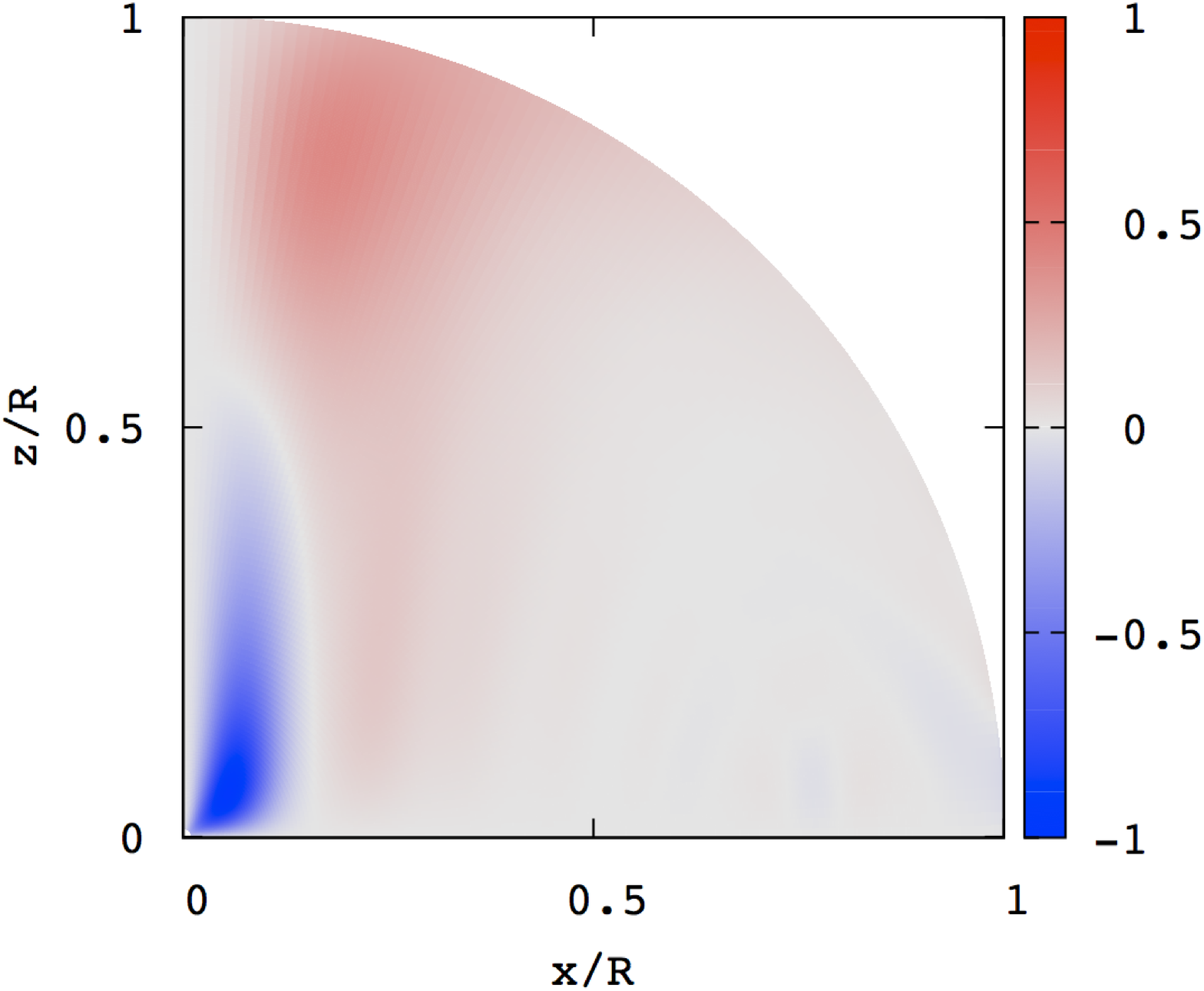}}
\resizebox{0.33\columnwidth}{!}{
\includegraphics{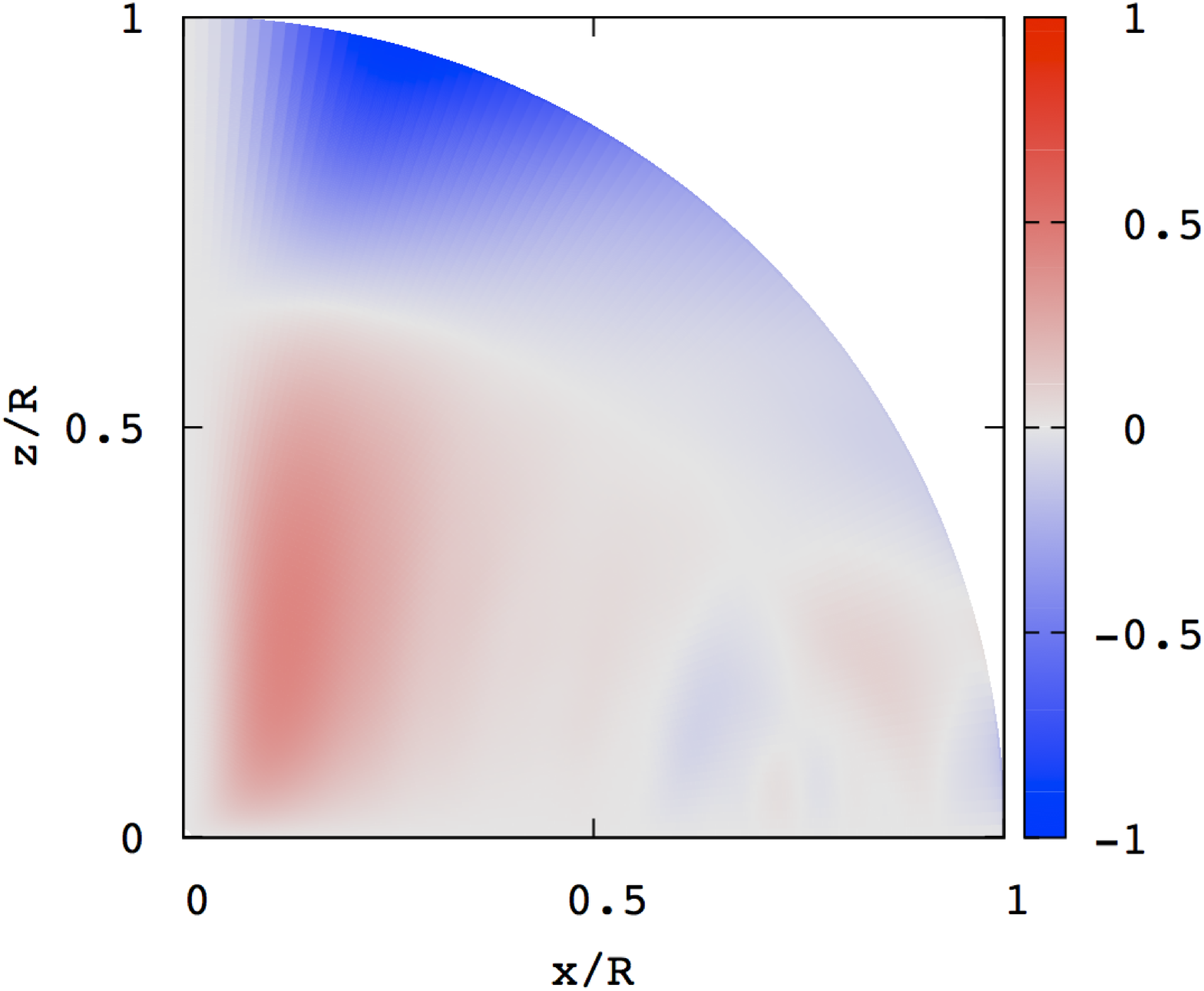}}
\end{center}
\caption{Same as Fig. 3 but for $\zeta_0=1$ where $\bar\omega=1.392\times10^{-2}$.
}
\label{fig:yievenz1}
\end{figure}

Figs. 5 and 6 show the patterns of $\hat\xi_j$ for the odd magnetic mode at $\zeta_0=0.01$ and $\zeta_0=1$,
where the frequency is $\bar\omega=5.846\times10^{-3}$ for the former and $\bar\omega=5.704\times10^{-3}$
for the latter.
The pattern $\hat\xi_r$ is antisymmetric and those of $\hat\xi_\theta$ and $\hat\xi_\phi$ are symmetric about
the equator for odd modes.
Interestingly, no significant differences in the wave patterns $\hat\xi_j$ are found between
the cases of $\zeta_0=0.01$ and $\zeta_0=1$ although the amplitudes of $\xi_\phi$ at $\zeta_0=1$ is
by about a factor 100 larger than those at $\zeta_0=0.01$.
The amplitudes of $\hat\xi_r$ and $\hat\xi_\theta$ are confined to the region along the magnetic axis, 
but $\hat\xi_\phi$ has amplitudes in the whole interior except on the magnetic axis.

\begin{figure}
\begin{center}
\resizebox{0.33\columnwidth}{!}{
\includegraphics{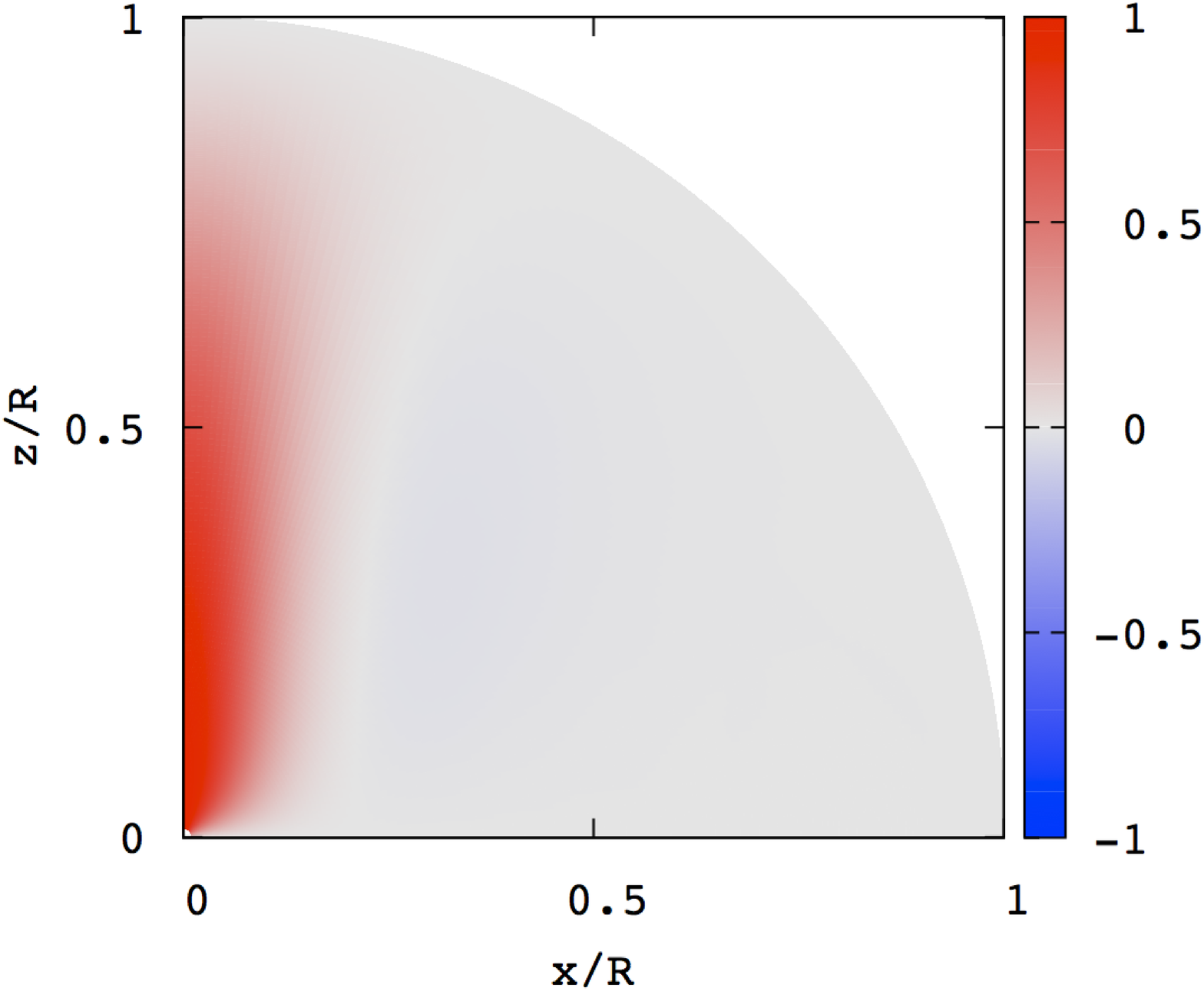}}
\resizebox{0.33\columnwidth}{!}{
\includegraphics{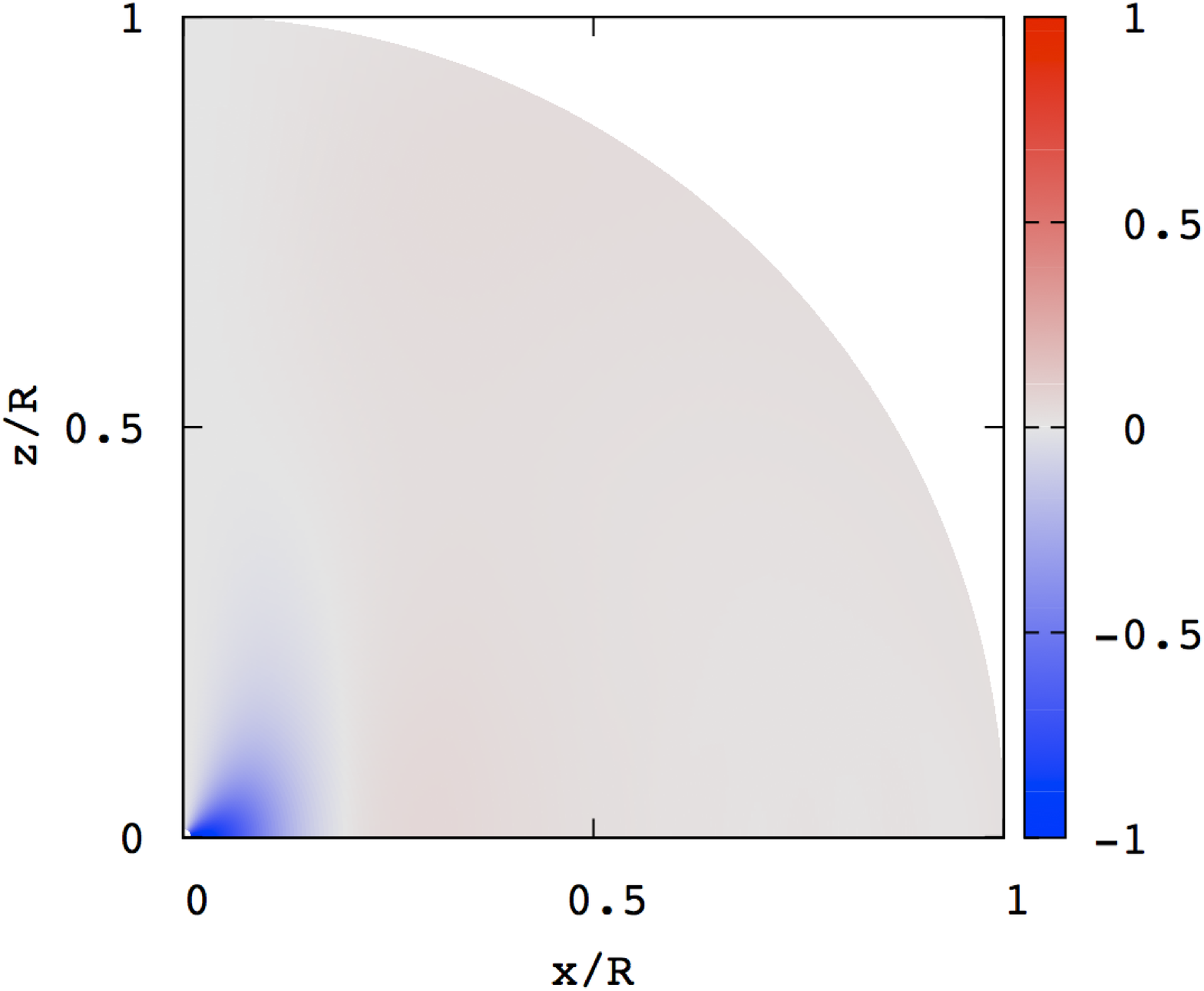}}
\resizebox{0.33\columnwidth}{!}{
\includegraphics{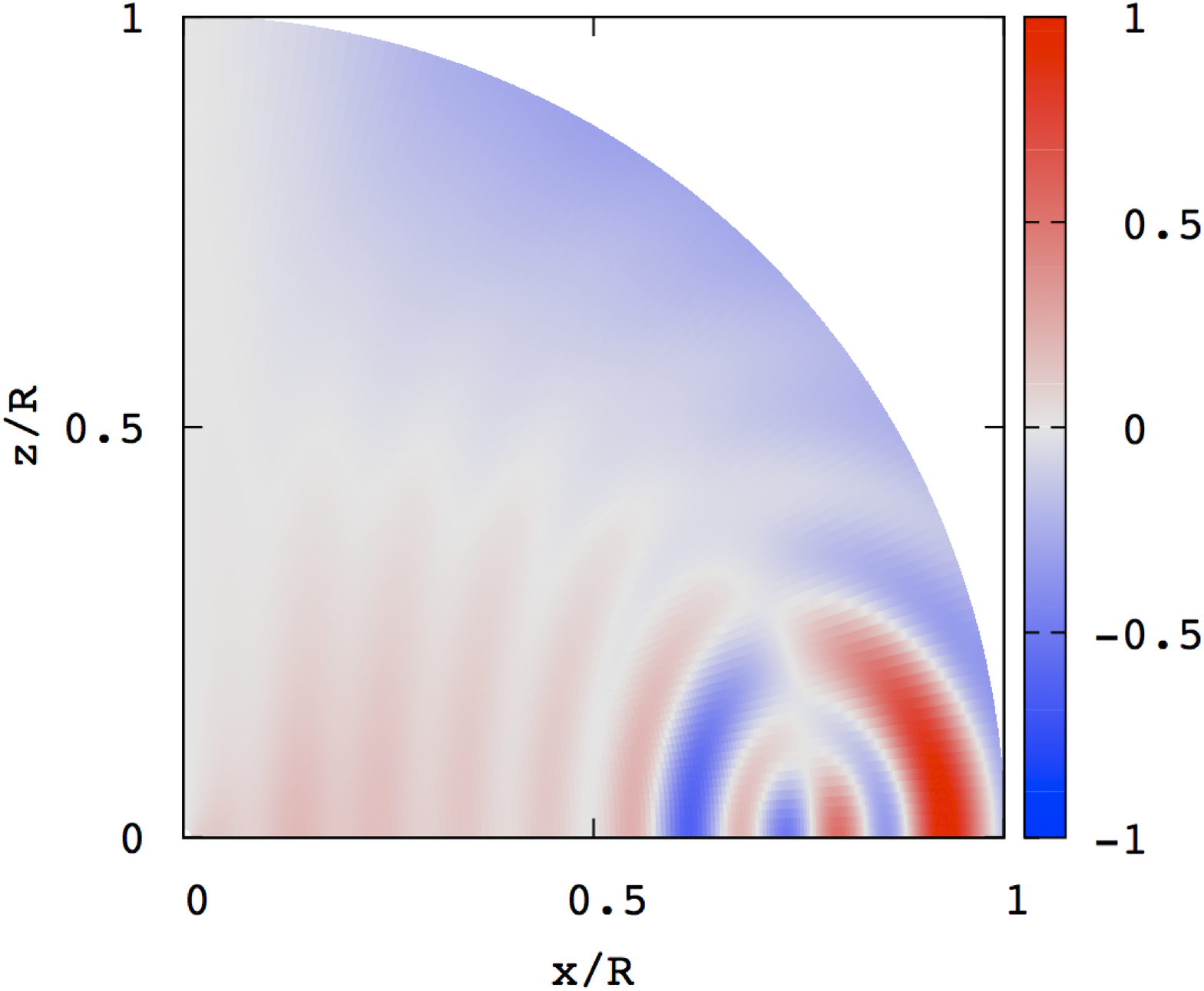}}
\end{center}
\caption{Wave patterns $\hat\xi_r$ (left), $\hat\xi_\theta$ (middle), and $\hat\xi_\phi$ (right) for the spheroidal magnetic mode of odd parity
at $\zeta_0=0.01$ where $\bar\omega=5.846\times10^{-3}$.
}
\label{fig:yioddz001}
\end{figure}

\begin{figure}
\begin{center}
\resizebox{0.33\columnwidth}{!}{
\includegraphics{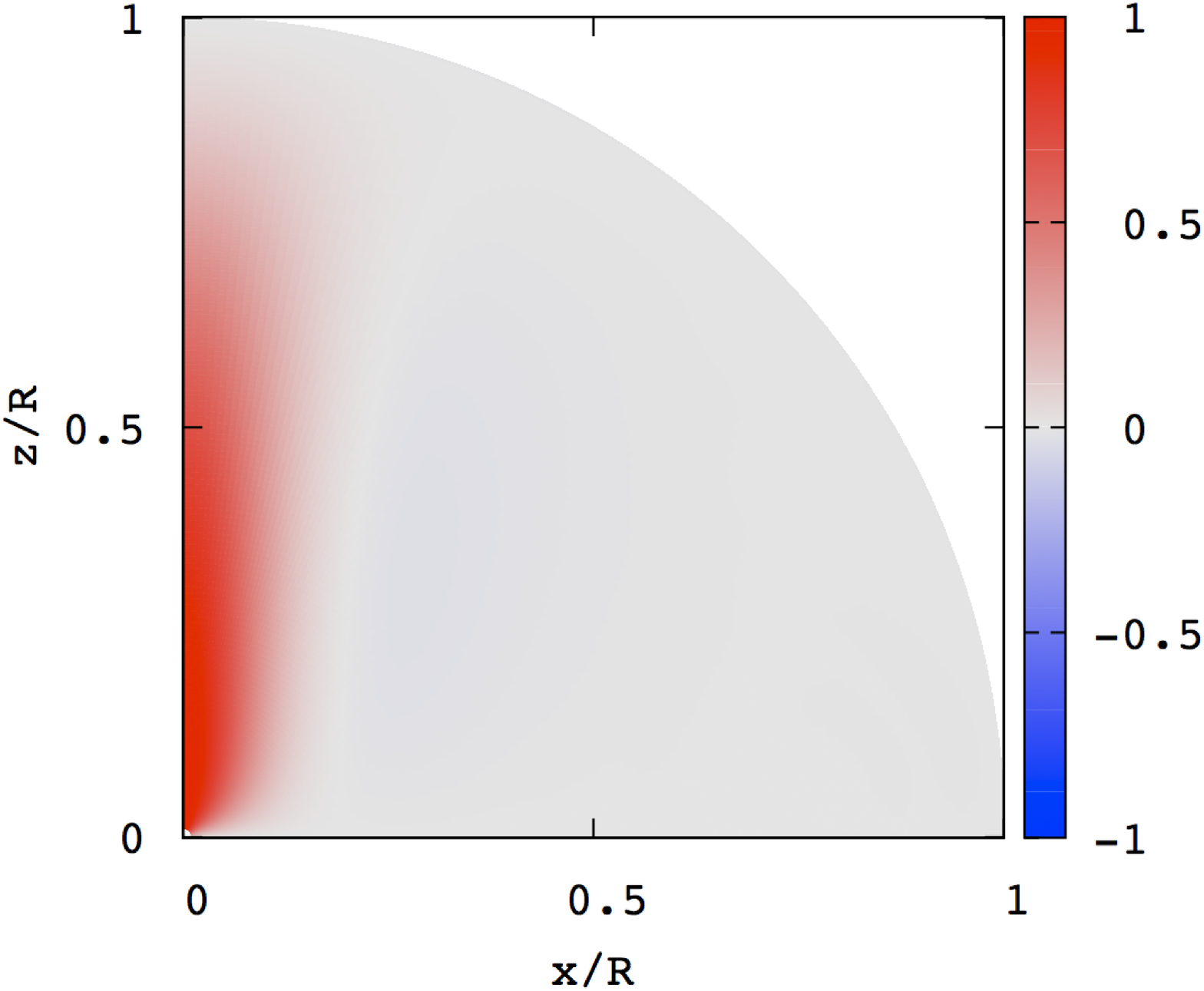}}
\resizebox{0.33\columnwidth}{!}{
\includegraphics{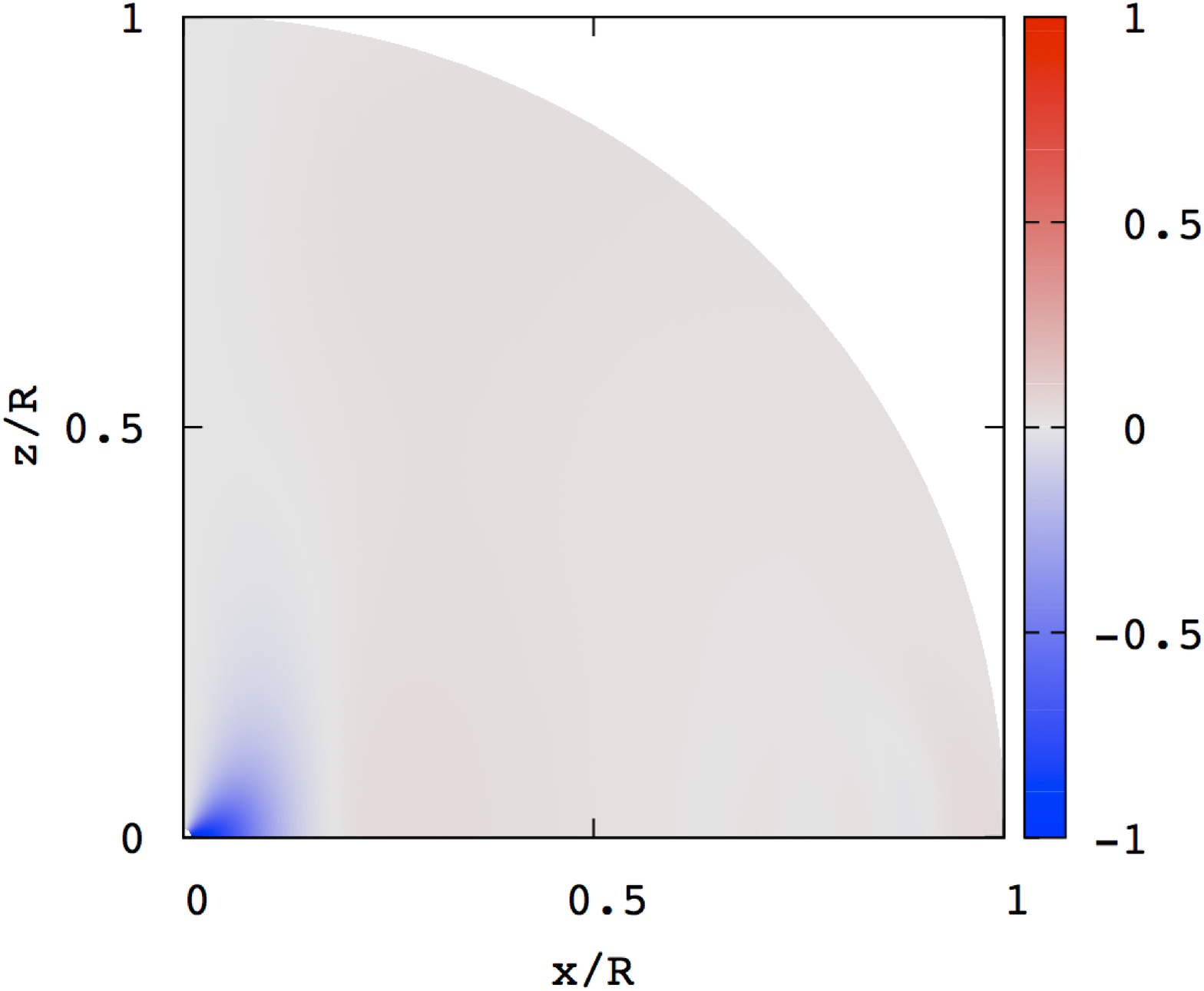}}
\resizebox{0.33\columnwidth}{!}{
\includegraphics{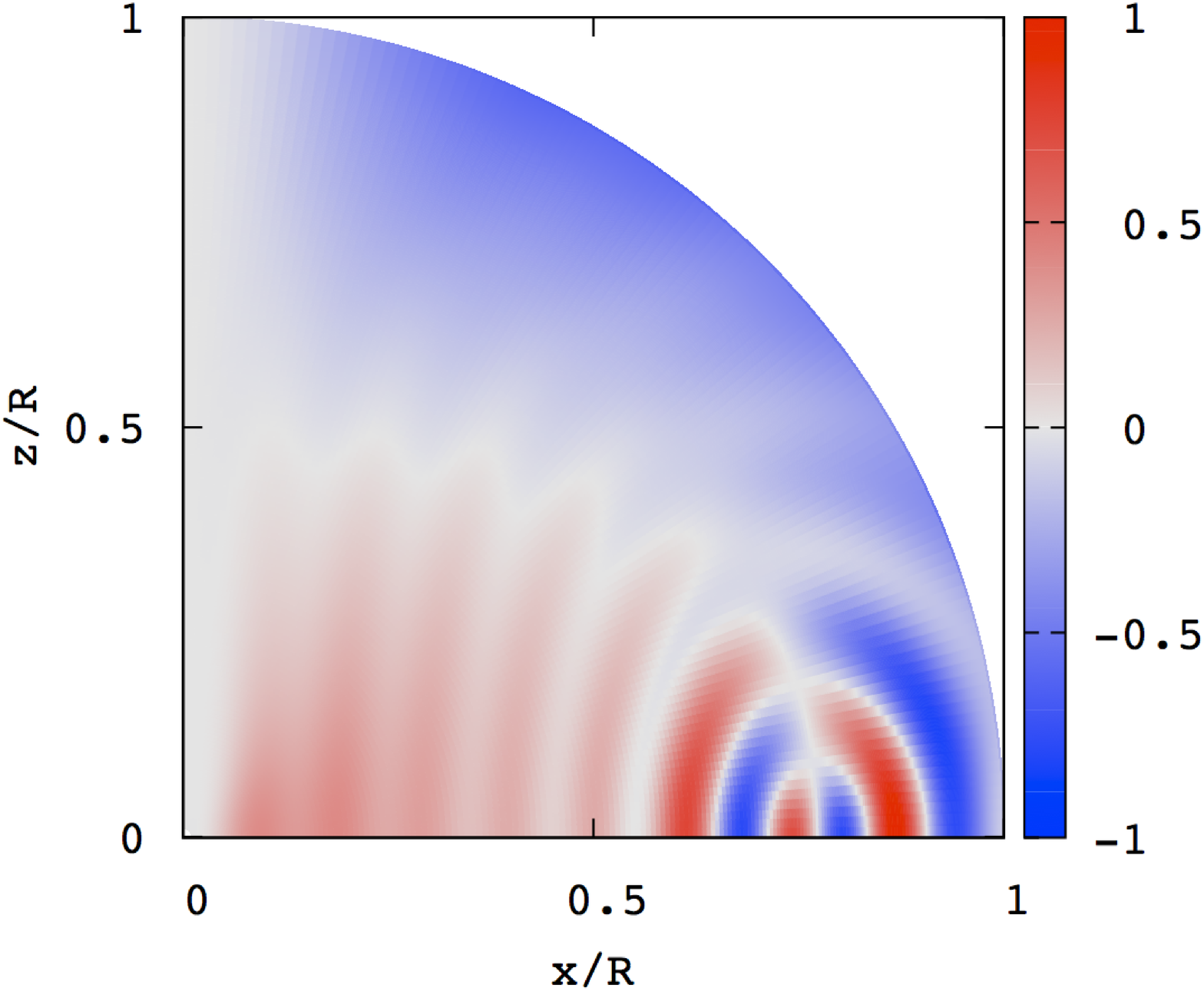}}
\end{center}
\caption{Same as Fig. \ref{fig:yioddz001} but for $\zeta_0=1$ where $\bar\omega=5.704\times10^{-3}$.
}
\label{fig:yioddz1}
\end{figure}

\subsection{Toroidal Modes}

\begin{figure}
\begin{center}
\resizebox{0.5\columnwidth}{!}{
\includegraphics{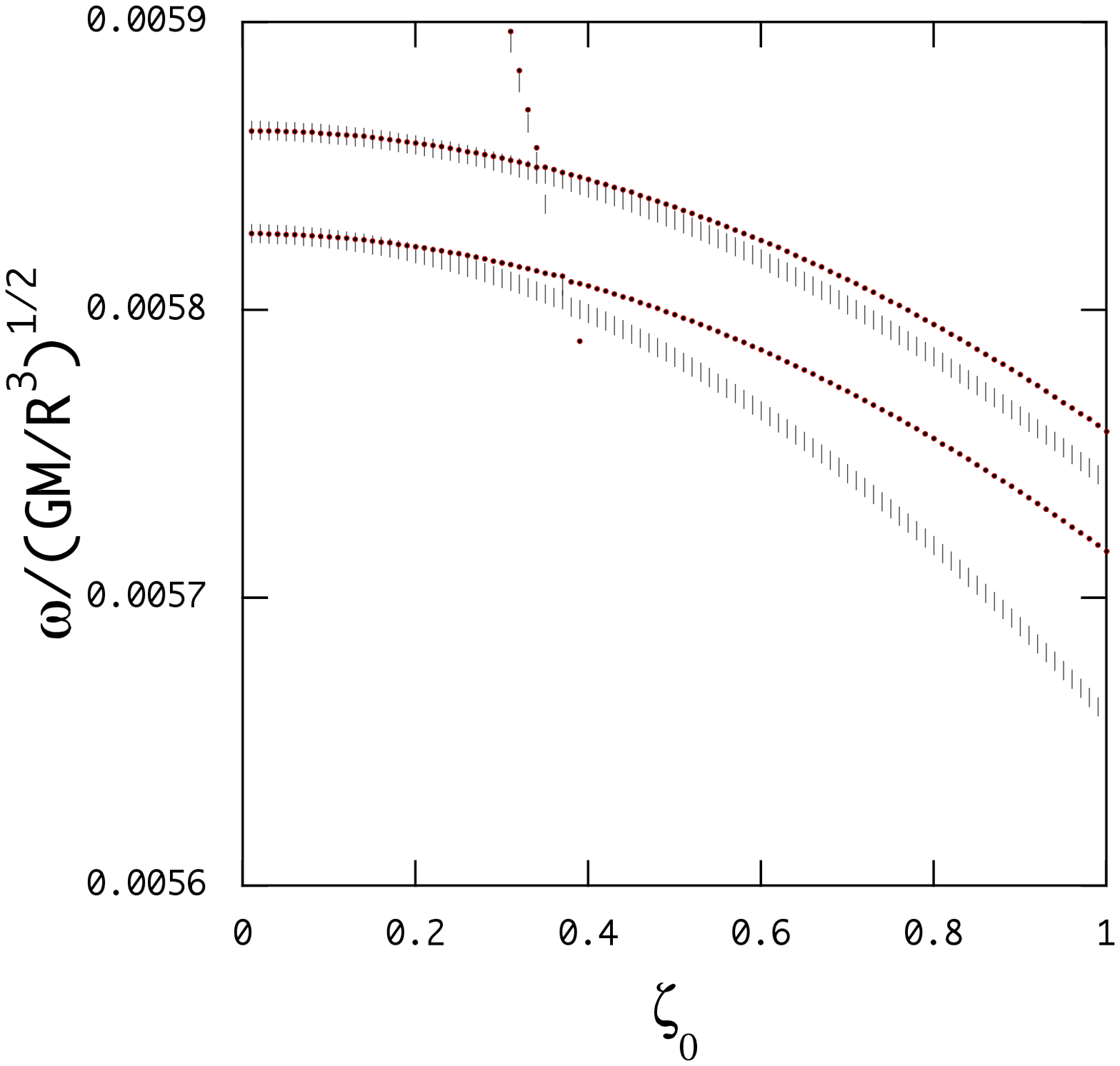}}
\end{center}
\caption{Eigenfrequency $\bar\omega$ of the first two toroidal magnetic modes of odd parity versus $\zeta_0$ for $B_S=10^{14}$G (red dots)
and $B_S=10^{15}$G (vertical short lines), where $10\times\bar\omega$ is plotted for $B_S=10^{14}$G.
}
\label{fig:yioddz1}
\end{figure}

In Fig. 7, $\bar\omega$ of the toroidal magnetic modes is plotted as a function of $\zeta_0$, 
where the red dots and short vertical lines are for the cases of
$B_S=10^{14}$G and $10^{15}$G, respectively, and $10\times\bar\omega$ is plotted for the former\footnote{At $\zeta_0=0$, axisymmetric toroidal modes and spheroidal modes are decoupled.
We have computed axisymmetric toroidal modes for $\zeta_0=0$ using the surface boundary conditions
$\pmb{y}_6=0$ and ${\rm d}\pmb{y}_5/{\rm d}r=0$, and 
the frequency $\bar\omega$ of the magnetic toroidal modes for $B_S=10^{15}$G is tabulated for the two boundary conditions in Table C1
in the Appendix C.
}.
The frequencies decrease with increasing $\zeta_0$, which behavior is similar to that found for spheroidal
magnetic modes, and we find $\delta_{\rm cs}\sim 10^{-3}$ for the toroidal modes except at and near the avoided crossings.
The frequencies $10\times\bar\omega$ for $B_S=10^{14}$G and $\bar\omega$ for $B_S=10^{15}$G
agree with each other only when $\zeta_0\ll 1$, and the agreement is lost
as $\zeta_0$ increases,
that is, except when $\zeta_0\sim 0$ the frequency of the toroidal magnetic modes does not linearly scale with
$B_S$, which is different from what we find for spheroidal magnetic modes.
The toroidal magnetic modes are governed by the differential equations (\ref{eq:y5}) and (\ref{eq:y6}) and
are described by the variables $\pmb{y}_5$ and $\pmb{y}_6$ when $\zeta_0=0$.
As $\zeta_0$ increases from $\zeta_0=0$, however,
the variables $\pmb{y}_1$ and $\pmb{y}_2$ come in and play a non-negligible role to determine the frequency
in equations (\ref{eq:y5}) and (\ref{eq:y6}).
Since the variables $\pmb{y}_1$ and $\pmb{y}_2$ are also affected by compressibility, 
which may be independent of the field strength $B$,
the proportionality of 
the frequency of the toroidal modes with the field strength $B_S$ may be lost.

Figures 8 and 9 show the wave patterns $\hat\xi_j$ for the toroidal modes at $\zeta_0=0.01$ and $\zeta_0=1$, respectively.
We find no significant differences in the wave patterns between the two cases, and the amplitudes of $\xi_\phi$
always dominate those of $\xi_r$ and $\xi_\theta$ for $\zeta_0\ltsim 1$.
These wave patterns may suggest that the eigenfunctions are not necessarily well converged for $j_{\rm max}=14$.

\begin{figure}
\begin{center}
\resizebox{0.33\columnwidth}{!}{
\includegraphics{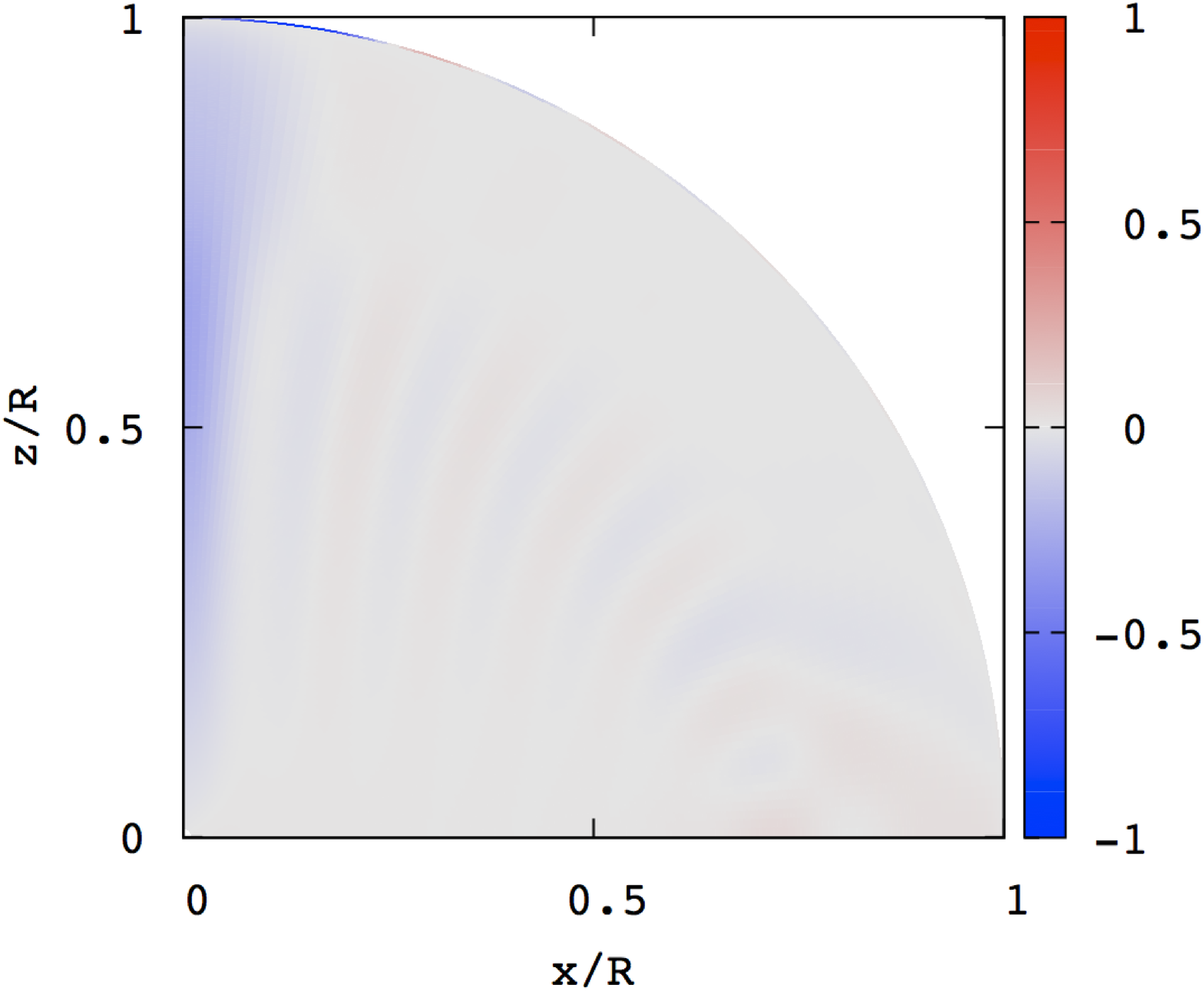}}
\resizebox{0.33\columnwidth}{!}{
\includegraphics{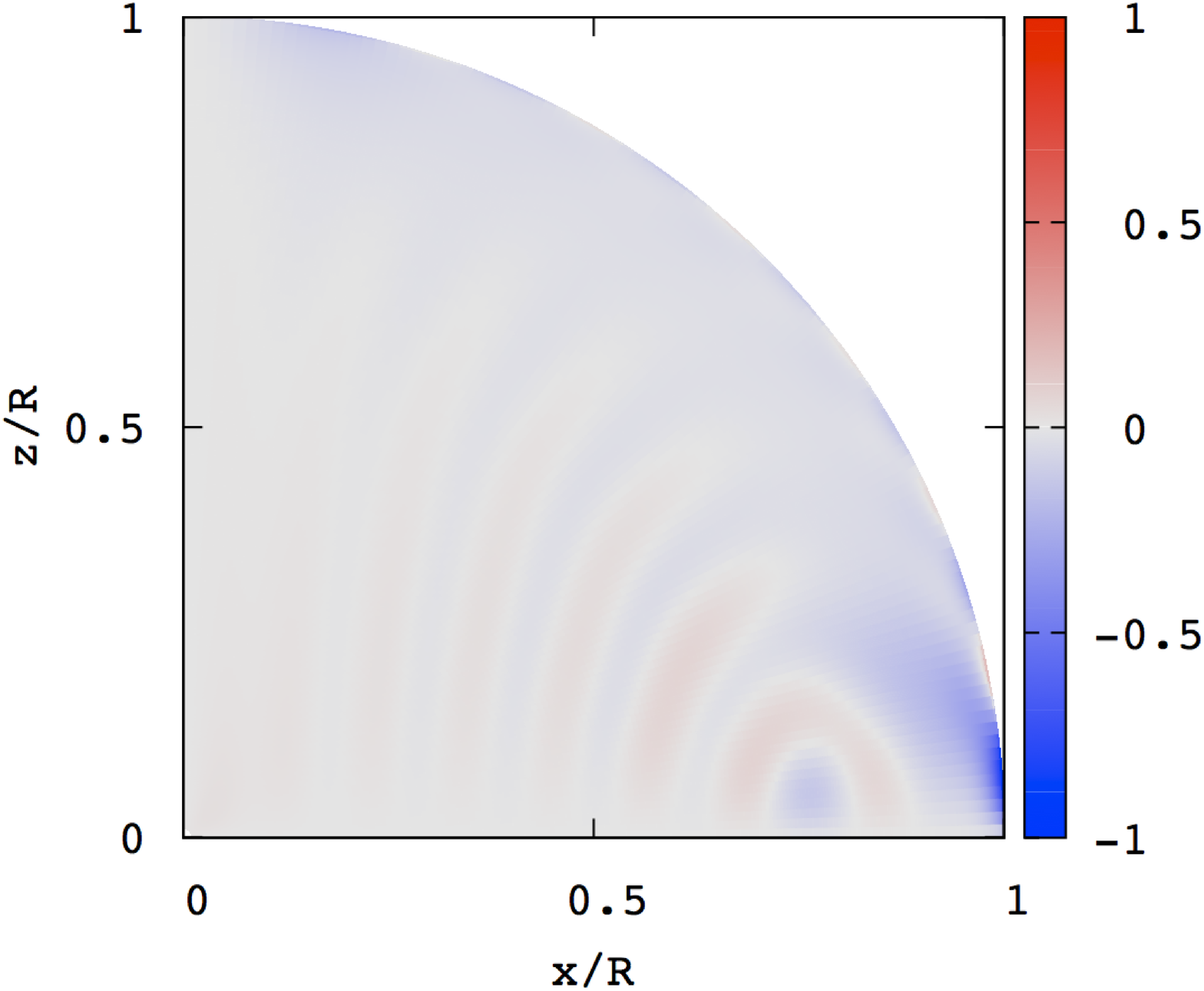}}
\resizebox{0.33\columnwidth}{!}{
\includegraphics{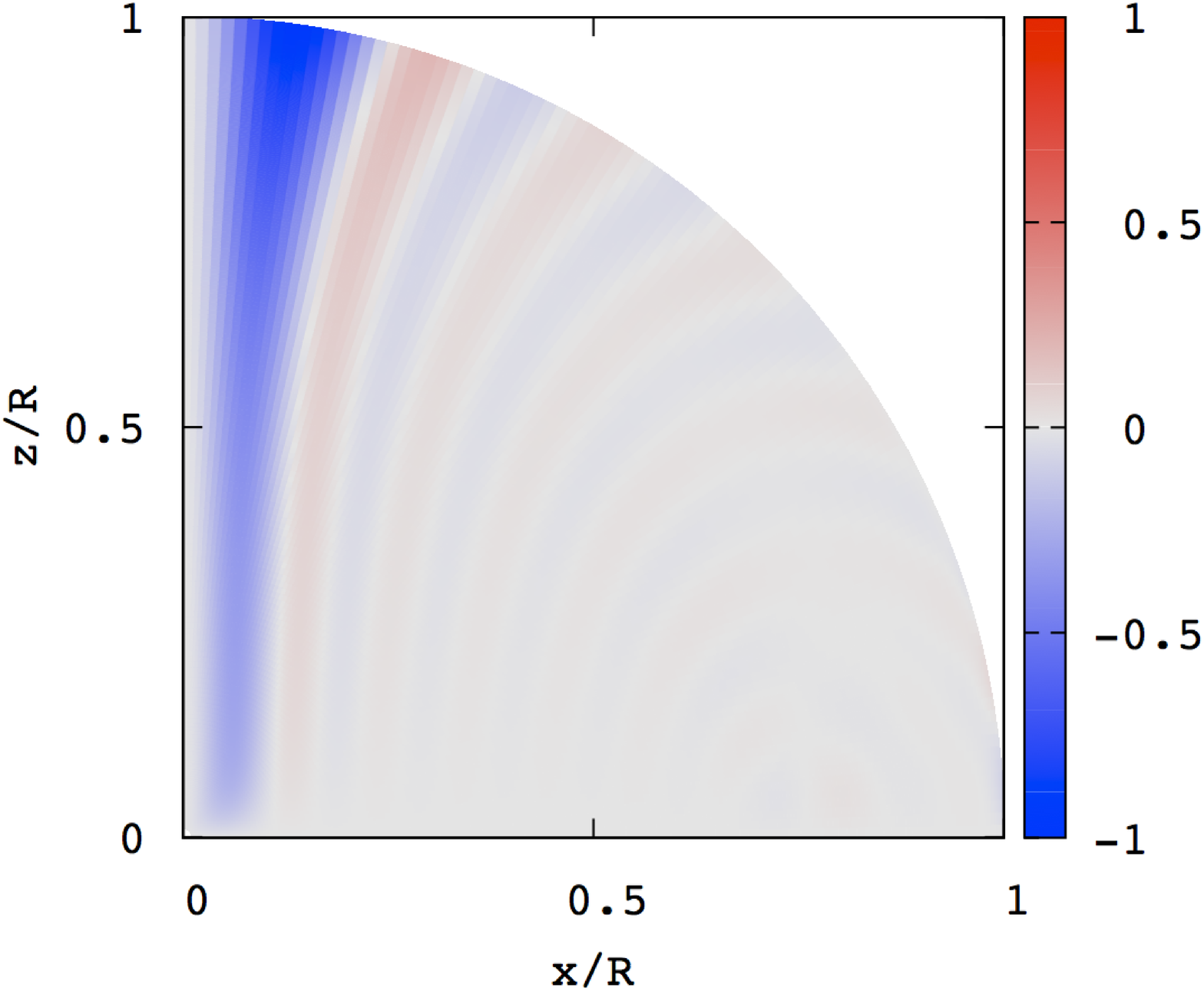}}
\end{center}
\caption{Wave patterns $\hat\xi_r$ (left), $\hat\xi_\theta$ (middle), and $\hat\xi_\phi$ (right) for the toroidal magnetic mode of odd parity
at $\zeta_0=0.01$ for $B_S=10^{15}$G where $\bar\omega=5.862\times10^{-3}$.
The mode tends to the pure toroidal magnetic mode of odd parity as $\zeta_0\rightarrow 0$.
}
\label{fig:yievenz001b15_toroid}
\end{figure}

\begin{figure}
\begin{center}
\resizebox{0.33\columnwidth}{!}{
\includegraphics{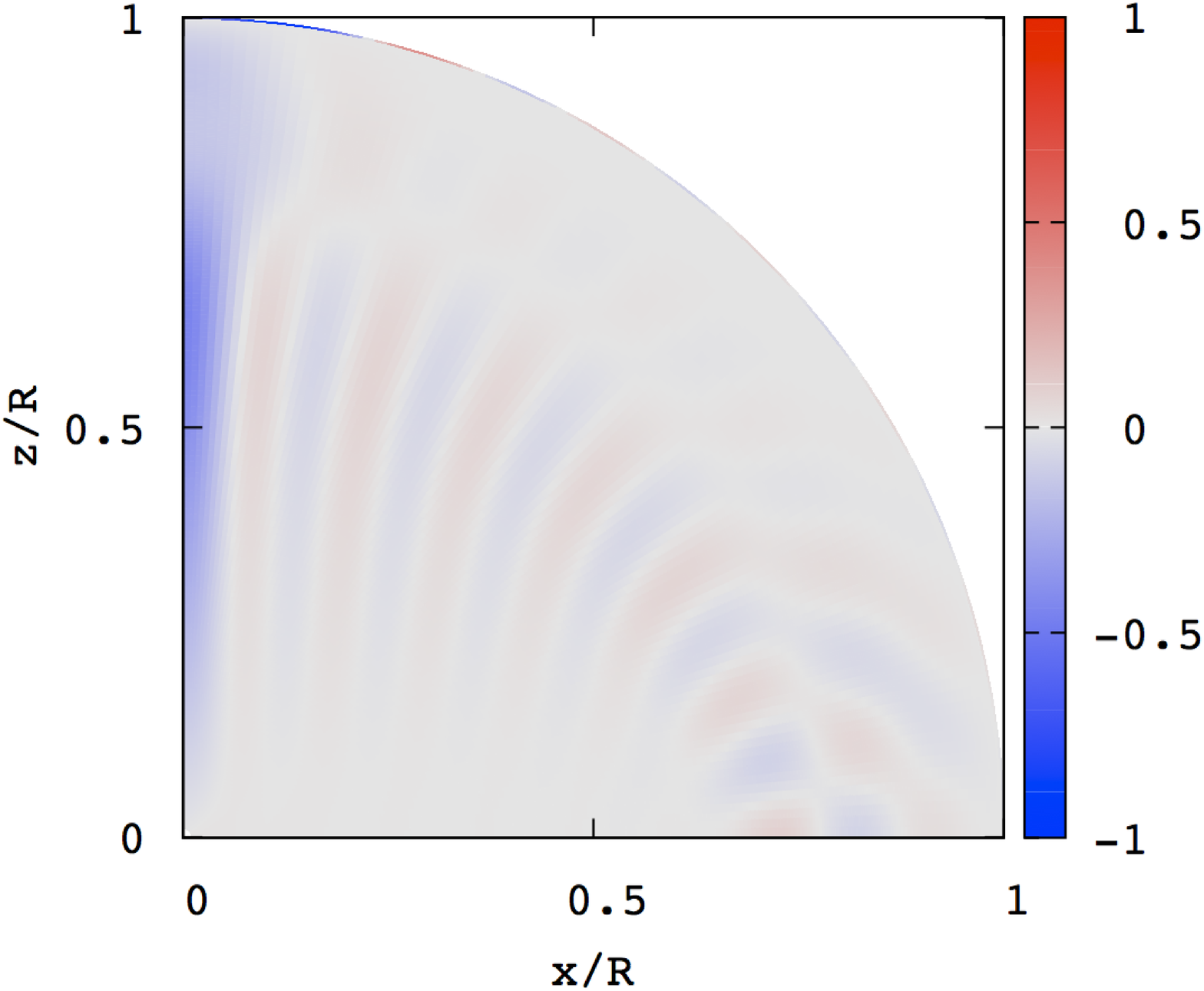}}
\resizebox{0.33\columnwidth}{!}{
\includegraphics{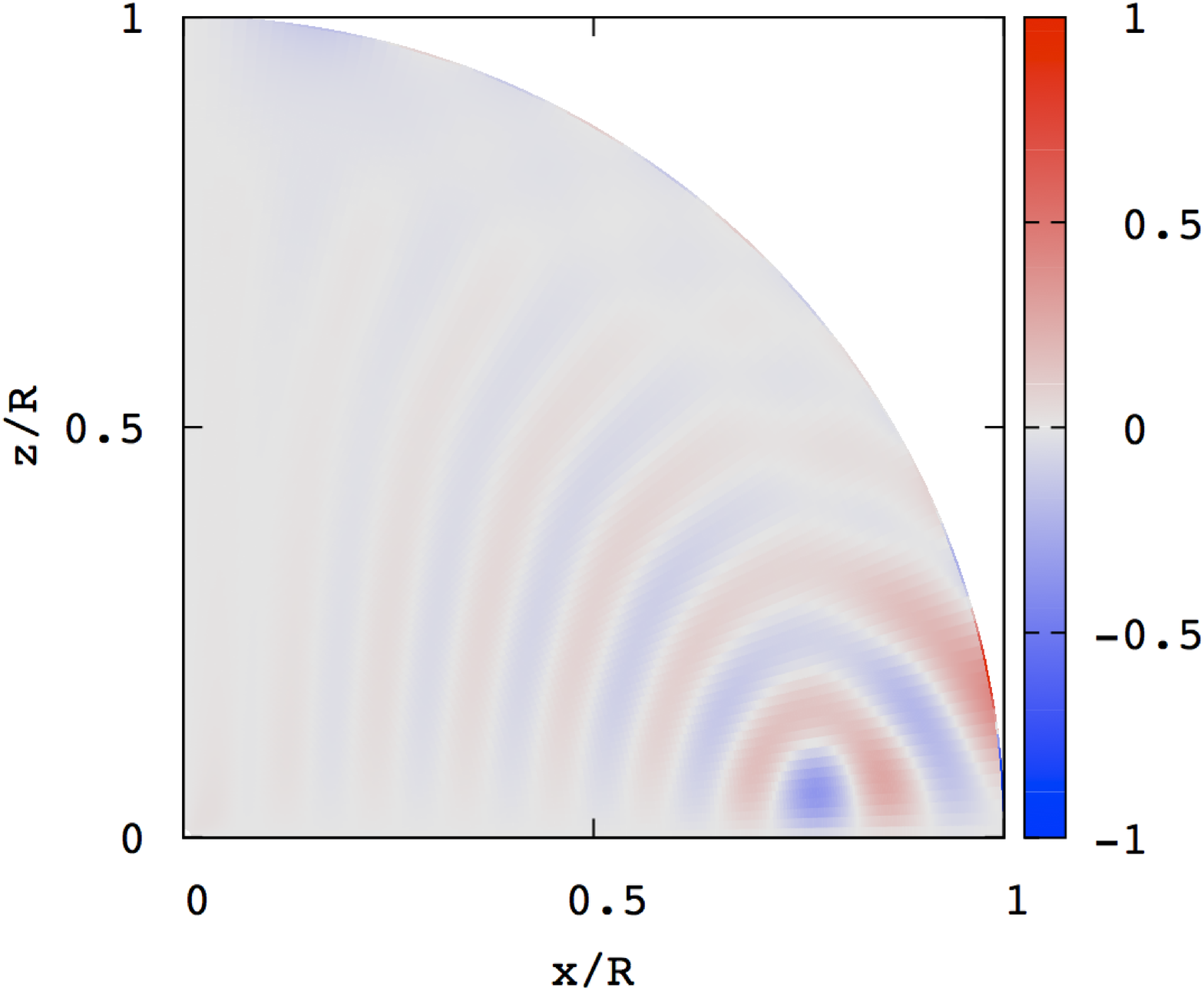}}
\resizebox{0.33\columnwidth}{!}{
\includegraphics{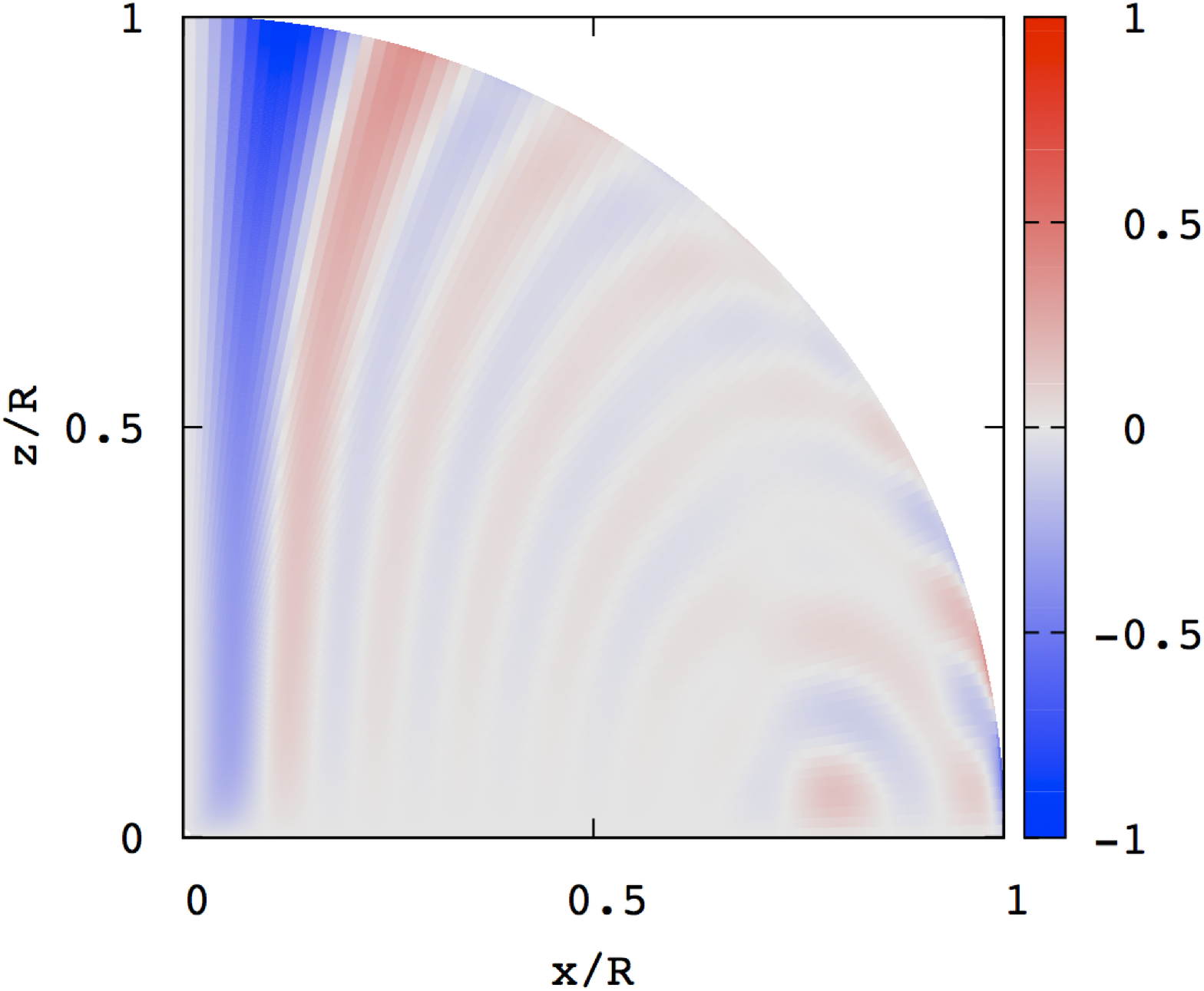}}
\end{center}
\caption{Same as Fig. \ref{fig:yievenz001b15_toroid} but for $\zeta_0=1$ where $\bar\omega=5.741\times10^{-3}$.
}
\label{fig:yievenz1b15_toroid}
\end{figure}

\section{Conclusion}

We have computed axisymmetric magnetic modes of neutron stars magnetized by a mixed poloidal and toroidal field.
For the mixed magnetic field ($\zeta_0\not=0$), axisymmetric spheroidal and toroidal modes are coupled.
Calculating magnetic modes as a function of $\zeta_0$ from $0$ to $\sim 1$, 
we find that the frequency decreases with increasing $\zeta_0$ and that they suffer avoided crossings
with other magnetic modes as $\zeta_0$ changes.
We also find that the frequency of the spheroidal magnetic modes is almost exactly
proportional to $B_S$ for $\zeta_0\ltsim 1$.
The amplitude of $\xi_\phi$ of the spheroidal magnetic modes is roughly proportional to $\zeta_0$
and become comparable to those of $\xi_r$ and $\xi_\theta$ when $\zeta_0\sim 1$.
For toroidal magnetic modes, on the other hand, the proportionality of the frequency to $B_S$ holds only when $\zeta_0\ll1 $ and is lost with increasing $\zeta_0$, and
$\xi_\phi$ is always dominating the other components for $\zeta_0\ltsim 1$.
We find no unstable modes of $\omega^2<0$ for axisymmetric magnetic modes.

We recognize several inconsistencies in our treatment of the oscillations of magnetized stars employed in this paper.
Firstly, we have ignored the equilibrium deformation due to magnetic fields.
The magnitudes of magnetic deformation may be of order of $B^2$, which is much smaller that the gas pressure in most of the 
interior region of the star, except for the case of extremely strong magnetic fields as strong as $B\sim10^{18}$G.
Secondly, the discontinuity in $B_\phi$ at the surface induces a surface current and hence the tangential force at the surface when $B_r\not=0$, which effects are not taken account of in our modal analyses.

As discussed in the Appendix C, the surface boundary conditions can be very important to determine 
the modal property of pure toroidal modes of a neutron star possessing a pure poloidal magnetic field.
For the surface boundary condition $B'_\phi=0$, we obtain only eigenmodes having real $\omega$ and eigenfunctions.
For the surface boundary condition ${\partial(\xi_\phi/r)/\partial r}=0$, however, we find an eigenmode having complex $\omega$ and eigenfunctions as well as real eigenmodes which have converging wave patterns with increasing $j_{\rm max}$.
For this boundary condition, the force operator $\pmb{F}(\pmb{\xi})$ is not self-adjoint and may permit complex modes.
The boundary condition ${\partial(\xi_\phi/r)/\partial r}=0$ may be used to satisfy the condition that the
traction vanishes at the surface of a solid crust. 
This may suggest that if we consider a neutron star with a solid crust the property of the toroidal magnetic modes in the fluid core
could be significantly different from what we have found in this paper.
Besides, if a neutron star has a solid crust,
there exist spheroidal and toroidal sound waves traveling in the solid crust.
It is interesting for us to examine how the sound waves in the solid crust will be affected by introducing
the toroidal magnetic field since the toroidal sound waves are possibly coupled with Alfv\'en modes in the core and could be quickly damped (e.g., Levin 2006, 2007).
Whether or not this strong damping of the crustal modes is still relevant for mixed poloidal and toroidal
field configurations is a question to be answered in terms of normal mode analyses
(e.g., Colaiuda \& Kokkotas 2012 in terms of MHD simulations).

Introducing the toroidal magnetic field, the spheroidal component of
the displacement vector of even parity is coupled with the toroidal component of odd parity and vice versa.
In the case of rotating stars, however, the Coriolis force couples the even (odd) spheroidal component with even (odd) toroidal component of the displacement vector.
This means that if we consider rotating stars magnetized by mixed poloidal and toroidal fields,
the perturbations are not separated into two different mode groups according to the parity, 
which probably makes it difficult to analyze the modal properties of
magnetized stars.

For mixed poloidal and toroidal magnetic field configurations, which are more favorable for magnetized stars
from the stability point of view,
toroidal (axial) and spheroidal (polar) components of the velocity fields of the perturbations
are coupled even for axisymmetric modes of non-rotating stars.
In this paper, we show that spheroidal and toroidal magnetic modes are separately obtained as discrete normal modes even with the coupling between them.
Because of the coupling, toroidal magnetic modes are accompanied by density perturbations, which
could affect the stability and hence the lifetime of the toroidal modes.
We have to take account of the effects of a solid crust on the magnetic modes to closely compare our results to those by Colaiuda \& Kokkotas (2012).
It is also interesting to examine whether or not the coherent magnetic modes discussed by Gabler et al (2016) and Passamonti \& Pons (2016) can be regarded as normal modes.
These will be among our future projects.

\appendix
\section{Equilibrium Magnetic Fields}

We consider neutron stars magnetized with a mixed poloidal and toroidal field.
We closely follow Colaiuda et al (2008) to derive the static mixed magnetic field in Newtonian gravity.
We start with a vector potential $\tilde{\pmb{A}}=(\tilde A_r, \tilde A_\theta,\tilde A_\phi)$
given in spherical polar coordinates $(r,\theta,\phi)$.
Applying a gauge transformation such that $A_\theta=\tilde A_\theta-\partial\chi/r\partial\theta=0$,
we will work with the gauged potential 
\be
\pmb{A}=(A_r,0,A_\phi),
\ee
which is assumed axisymmetric so that $\partial\pmb{A}/\partial\phi=0$.
With this vector potential, magnetic fields may be given by
\be
\pmb{B}=\nabla\times\pmb{A}=\pmb{e}_r{1\over r\sin\theta}{\partial\over\partial\theta}\sin\theta A_\phi
+\pmb{e}_\theta\left(-{1\over r}{\partial rA_\phi\over \partial r}\right)
+\pmb{e}_\phi\left(-{1\over r}{\partial A_r\over\partial \theta}\right),
\ee
and the Amp\'ere law $\nabla\times\pmb{B}=4\pi\pmb{J}/c$ reduces to
\begin{eqnarray}
{4\pi\over c}\pmb{J}
=-\left({1\over r^2\sin\theta}{\partial\over\partial\theta}\sin\theta{\partial\over\partial\theta}A_r\right)\pmb{e}_r
+\left({1\over r}{\partial^2\over\partial r\partial\theta}A_r\right)\pmb{e}_\theta
-\left(\nabla^2A_\phi-{1\over r^2\sin^2\theta}A_\phi\right)\pmb{e}_\phi,
\end{eqnarray}
where $\pmb{J}$ is the current density, 
and $c$ denotes the velocity of light.
For the Lorenz force we may write
\begin{eqnarray}
\left(\nabla\times\pmb{B}\right)\times\pmb{B}
&=&-\pmb{e}_r\left[{1\over r}{\partial rA_\phi\over\partial r}\left(\nabla^2A_\phi-{1\over r^2\sin^2\theta}A_\phi\right)+{1\over r}{\partial A_r\over\partial\theta}{1\over r}{\partial^2A_r\over\partial r\partial\theta}\right]\nonumber\\
&&-\pmb{e}_\theta\left[{1\over r}{\partial A_r\over \partial\theta}{1\over r^2\sin\theta}{\partial\over\partial\theta}\sin\theta{\partial\over\partial\theta}A_r+\left({1\over r\sin\theta}{\partial\over\partial\theta}\sin\theta A_\phi\right)\left(\nabla^2A_\phi-{1\over r^2\sin^2\theta}A_\phi\right)\right]\nonumber\\
&&+\pmb{e}_\phi\left[{1\over r}{\partial rA_\phi\over\partial r}{1\over r^2\sin\theta}{\partial\over\partial\theta}
\sin\theta{\partial\over\partial\theta}A_r-\left({1\over r\sin\theta}{\partial\over\partial\theta}\sin\theta A_\phi\right)\left({1\over r}{\partial^2A_r\over\partial r\partial\theta}\right)\right].
\end{eqnarray}

Assuming that the $\phi$ component of the Lorenz force vanishes identically, we obtain
\begin{eqnarray}
{1\over r^3\sin^2\theta}\left[\left({\partial \over\partial r}r\sin\theta A_\phi\right)\left({\partial\over\partial\theta}
\sin\theta{\partial\over\partial\theta}A_r\right)-\left({\partial\over\partial\theta}r\sin\theta A_\phi\right)\left({\partial\over\partial r}\sin\theta{\partial\over\partial\theta}A_r\right)\right]=0,
\end{eqnarray}
which is equivalent to, when $\partial A_j/\partial\phi=0$,
\be
\nabla \left(r\sin\theta A_\phi\right)\times\nabla\left(\sin\theta{\partial\over\partial\theta}A_r\right)=0.
\ee
We therefore assume, using an arbitrary function $H(x)$,
\be
\sin\theta{\partial A_r\over\partial\theta}=H(r\sin\theta A_\phi).
\ee
For the function $H(x)$, we employ $H(x)=\zeta x$ so that
\be
{\partial\over\partial\theta}A_r=\zeta r A_\phi,
\ee
where $\zeta$ is a constant.
Introducing a function $a(r,\theta)$ that satisfies
\be
A_\phi={1\over r}{\partial a\over \partial\theta}, \quad A_r=\zeta a,
\ee
we obtain
\be
\pmb{B}={1\over r^2\sin\theta}{\partial\over\partial\theta}\sin\theta{\partial a\over\partial\theta}\pmb{e}_r
-{1\over r}{\partial^2a\over\partial r\partial\theta}\pmb{e}_\theta
-\zeta{1\over r}{\partial a\over\partial\theta}\pmb{e}_\phi,
\ee
\begin{eqnarray}
{4\pi\over c}\pmb{J}
=-\zeta B_r\pmb{e}_r-\zeta B_\theta\pmb{e}_\theta
-\left[\nabla^2\left({1\over r}{\partial a\over\partial\theta}\right)-{1\over r^2\sin^2\theta}
{1\over r}{\partial a\over\partial\theta}\right]\pmb{e}_\phi,
\end{eqnarray}
and
\be
\left(\nabla\times\pmb{B}\right)\times\pmb{B}
={4\pi\over c}\hat J_\phi
{1\over r\sin\theta}\nabla\left(\sin\theta{\partial a\over\partial\theta}\right),
\ee
where
\be
{4\pi\over c}\hat J_\phi={4\pi\over c}J_\phi+\zeta B_\phi.
\ee

The equilibrium state of a magnetized star is given by
\be
{1\over\rho}\nabla p+\nabla\Phi-{1\over 4\pi\rho}\left(\nabla\times\pmb{B}\right)\times\pmb{B}=0,
\ee
which may reduce to
\be
\nabla\left(u+{p\over\rho}+\Phi\right)={1\over \rho c}{\hat J_\phi\over r\sin\theta}\nabla\left(\sin\theta{\partial a\over\partial\theta}\right),
\ee
where we have used the first law of thermodynamics for adiabatic flows given by
\be
du=-pd\left({1\over \rho}\right),
\ee
and $u$ denotes the internal energy per unit mass.
Since
\be
0=\nabla\times\nabla\left(u+{p\over\rho}+\Phi\right)=\nabla\left({1\over \rho c}{\hat J_\phi\over r\sin\theta}\right)\times\nabla\left(\sin\theta{\partial a\over\partial\theta}\right),
\ee
we may write, using an arbitrary function $F(x)$,
\be
{1\over \rho c}{\hat J_\phi\over r\sin\theta}=F\left(\sin\theta{\partial a\over\partial\theta}\right).
\ee
Since $a\propto B\equiv |\pmb{B}|$, the function $F(x)$ may be expanded in terms of the field strength $B$ as
\be
F\left(\sin\theta{\partial a\over\partial\theta}\right)=c_0+c_1\sin\theta{\partial a\over\partial\theta}+O\left(B^2\right),
\ee
where $c_0$ and $c_1$ are expansion constants.
Keeping only the first term $c_0$, we obtain
\be
4\pi c_0\rho  r\sin\theta={4\pi\over c}\hat J_\phi=-\left[\nabla^2\left({1\over r}{\partial a\over\partial\theta}\right)-{1\over r^2\sin^2\theta}
{1\over r}{\partial a\over\partial\theta}\right]-\zeta^2{1\over r}{\partial a\over\partial\theta}.
\label{eq:gradshaf}
\ee
When we ignore the equilibrium deformation due to magnetic fields so that $\rho$ depends only on the radial distance $r$ from the centre,
substituting into (\ref{eq:gradshaf}) for simplicity
\be
a(r,\theta)=a_1(r)P_1(\cos\theta)
\label{eq:a1p1}
\ee
with $P_1(\cos\theta)=\cos\theta$ being a Legendre polynomial, we obtain
\be
{d^2a_1\over dr^2}+\left(\zeta^2-{2\over r}\right)a_1=4\pi c_0\rho r^2,
\ee
which may be rewritten as
\be
{d^2f\over dr^2}+{4\over r}{df\over dr}+\zeta^2 f=-4\pi c_0\rho,
\label{eq:feq}
\ee
where
\be
f=-{a_1\over r^2},
\ee
and 
$
f(r)=\alpha_0+O(r^2)
$
is assumed at the stellar centre with a constant $\alpha_0$.
Numerically integrating equation (\ref{eq:feq}) from the stellar centre to the surface, we may 
determine the two constants $c_0$ and $\alpha_0$ so that the two boundary conditions are satisfied 
at the stellar surface.
Using the function $f$ thus computed, we write
\be
B_r=2f(r)\cos\theta, \quad B_\theta=-{1\over r}{dr^2f(r)\over dr}\sin\theta\equiv {\cal Q}\sin\theta, \quad B_\phi=-\zeta rf(r)\sin\theta,
\ee
and $\zeta$ may be treated as a free parameter.

\section{Surface Boundary Conditions}

The surface boundary conditions we apply at $r=R$ are
\be
\left[\delta\pmb{B}\right]^+_-=0,
\quad \left[\delta\left(p+{1\over 8\pi}\pmb{B}^2\right)\right]^+_-=0, 
\ee
where 
\be
\left[f(r)\right]^+_-=\lim_{\epsilon\rightarrow 0}\left[f(r+\epsilon)-f(r-\epsilon)\right].
\ee
Since we have assumed the continuity of the functions $f$ and $df/dr$ to the functions $f^{\rm ex}=\mu_b/r^3$ and $df^{\rm ex}/dr$
at the surface, we have at the surface
\be
\left[B_r\right]^+_-=0, \quad \left[B_\theta\right]^+_-=0.
\ee
But, since we assume the parameter $\zeta\not=0$ for $r<R$ and $\zeta=0$ for $r>R$,
we have at $r=R$
\be
\left[B_\phi(R)\right]^+_-=\zeta Rf\sin\theta\not=0,
\ee
which suggests the existence of the surface current given by 
\be
J_\theta=\zeta{c\over 4\pi}{a\over r}\sin\theta\delta(r-R),
\ee
where $\delta(r-R)$ is the delta function.
The existence of the surface current, however, induces a tangential force given by $J_\theta\pmb{e}_\theta\times\pmb{B}$
unless $\pmb{n}\cdot\pmb{B}=0$ at the surface, where $\pmb{n}$ denotes the unit normal vector to the surface (e.g., Melrose 1986).
Since $\pmb{n}\cdot\pmb{B}\not=0$ is assumed in this paper, we have $J_\theta\pmb{e}_\theta\times\pmb{B}\not=0$, which may suggest
the inconsistency will be serious for large values of $\zeta$.

Neglecting the displacement current in the perturbed Ampere law we get $\nabla\times\pmb{B}^{\prime (+)}=0$ outside the star, 
where we have assumed $p=\rho=0$ and $\pmb{J}=0$.
For this $\pmb{B}^{\prime (+)}$ we may write $\pmb{B}^{\prime (+)}=\nabla\psi$, where
we have used the superscript $^{(+)}$ to indicate quantities for $r>R$ and
we shall use a superscript $^{(-)}$ for $r<R$.
Since $\nabla\cdot\pmb{B}'=0$,
the scalar potential $\psi$ satisfies the Laplace equation $\nabla^2\psi=0$, the solution of which may be given as
\be
\psi=\sum_{l=0}^\infty c_lr^{-l-1}Y_l^0(\theta,\phi)e^{\rmi\omega t},
\ee
where $c_l$ is the expansion coefficients 
and we have assumed that the potential $\psi$ is axisymmetric ($m=0$) and vanishes at infinity.
We then obtain
\be
B'^{(+)}_r=-\sum_l(l+1)c_lr^{-l-2}Y_l^0(\theta,\phi)e^{\rmi\omega t}, 
\quad B'^{(+)}_\theta=\sum_lc_lr^{-l-2}{\partial\over\partial\theta}Y_l^0(\theta,\phi)e^{\rmi\omega t}, 
\quad B'^{(+)}_\phi=0.
\label{eq:outermag}
\ee

The surface condition $\left[\delta\pmb{B}\right]^+_-=0$ may lead to
\be
\left[B'_r-{B_\phi\over r}\xi_\phi\right]^+_-=B'^{(+)}_r
-\left(B'^{(-)}_r+{\zeta f\sin\theta}\xi_\phi^{(-)}\right)=0,
\label{eq:deltabr}
\ee
\be
\left[B'_\theta+\xi_r{\partial B_\theta\over \partial r}-\xi_\phi {B_\phi\over r}{\cos\theta\over\sin\theta}\right]^+_-=B'^{(+)}_\theta
-\left(B'^{(-)}_\theta+\left[{d{\cal Q}\over d r}\right]^+_-\sin\theta\xi_r^{(-)}
+ \zeta f {\cos\theta}\xi_\phi^{(-)}\right)=0,
\label{eq:deltabtheta}
\ee
\be
\left[B'_\phi+\xi_r{\partial B_\phi\over \partial r}+\xi_\theta{\partial\over\partial\theta}{B_\phi\over r}\right]^+_-=B'^{(+)}_\phi-\left(B'^{(-)}_\phi-\zeta {drf \over d r}\sin\theta \xi_r^{(-)}
-\zeta f\cos\theta \xi_\theta^{(-)}\right)=0,
\label{eq:deltabphi}
\ee
and the mechanical force balance condition at the surface to
\begin{eqnarray}
0=\left[\delta\left(p+{1\over 8\pi}\pmb{B}^2\right)\right]^+_-&=&\left[\delta p\right]^+_-+{1\over 4\pi}\left[\pmb{B}\cdot\delta\pmb{B}\right]^+_-\nonumber\\
&=&-\delta p^{(-)}+{1\over 4\pi}\zeta Rf\sin\theta\left(B'^{(-)}_\phi-\zeta {d rf \over d r}\sin\theta\xi_r^{(-)}
-\zeta f\cos\theta\xi_\theta^{(-)}+\left({\cal Q} +2f\right)\cos\theta{\xi_\phi^{(-)}\over r}\right).
\label{eq:mechbc}
\end{eqnarray}

If we write
\be
{B_r'^{(+)}\over f}=\sum_jc_{l'_j}^SY_{l'_j}^0(\theta,\phi)e^{\rmi\omega t}, 
\quad {B_\theta'^{(+)}\over f}=\sum_jc_{l'_j}^H{\partial\over\partial\theta}Y_{l'_j}^0(\theta,\phi)e^{\rmi\omega t}, 
\quad {B_\phi'^{(+)}\over f}=-\sum_jc_{l'_j}^T{\partial\over\partial\theta}Y_{l'_j}^0(\theta,\phi)e^{\rmi\omega t},
\ee
equations (\ref{eq:outermag}) at the surface reduce to
\be
\pmb{c}^S+\pmbmt{L}^+\pmb{c}^H=0, \quad \pmb{c}^T=0,
\label{eq:csch}
\ee
where
\be
\pmb{c}^S=\left(c^S_{l'_j}\right), \quad \pmb{c}^H=\left(c^H_{l'_j}\right), \quad \pmb{c}^T=\left(c^T_{l'_j}\right),
\ee
and
\be
\left(\pmbmt{L}^+\right)_{ij}=\left(l'_j+1\right)\delta_{ij}.
\ee
The boundary conditions (\ref{eq:deltabr}), (\ref{eq:deltabtheta}), and (\ref{eq:deltabphi}) can be rewritten as
\be
\pmb{c}^S=\pmb{b}^S+\zeta_0\pmbmt{C}_1\pmb{y}_5,
\ee
\be
\pmbmt{C}_0\pmb{c}^H=\pmbmt{C}_0\pmb{b}^H+\zeta_0^2\left(\pmbmt{1}-\pmbmt{Q}_0\pmbmt{Q}_1\right)\pmb{y}_1
-\zeta_0\pmbmt{Q}_0\pmbmt{C}_1\pmb{y}_5,
\ee
\be
\pmbmt{C}_0\pmb{c}^T=\pmbmt{C}_0\pmb{b}^T-\zeta_0\left({{\cal Q}\over f}+1\right)\left(\pmbmt{1}-\pmbmt{Q}_0\pmbmt{Q}_1\right)\pmb{y}_1+\zeta_0\pmbmt{Q}_0\pmbmt{C}_1\pmb{y}_3=0,
\ee
and the condition (\ref{eq:mechbc}) as
\begin{eqnarray}
\pmb{y}_2-\pmb{y}_1
=-\zeta_0\beta\left({{\cal Q}\over f}+2\right)\pmbmt{Q}_0\pmbmt{C}_1\pmb{y}_5,
\end{eqnarray}
where $\zeta_0\equiv \zeta R$ and we have used at the surface
\be
\left[{d\ln{\cal Q}\over d\ln r}\right]^+_-=\zeta_0^2{f\over {\cal Q}}.
\ee
Eliminating the vectors $\pmb{c}^S$ and $\pmb{c}^H$ using equation (\ref{eq:csch}), we obtain
\be
\pmb{b}^S+\zeta_0\pmbmt{C}_1\pmb{y}_5+\pmbmt{L}^+\left[\pmb{b}^H+\zeta_0^2\pmbmt{C}_0^{-1}\left(\pmbmt{1}-\pmbmt{Q}_0\pmbmt{Q}_1\right)\pmb{y}_1-\zeta_0\pmbmt{C}_0^{-1}\pmbmt{Q}_0\pmbmt{C}_1\pmb{y}_5\right]=0,
\ee
and
\be
\pmb{b}^T=\zeta_0\pmbmt{C}_0^{-1}\left[\left({{\cal Q}\over f}+1\right)\left(\pmbmt{1}-\pmbmt{Q}_0\pmbmt{Q}_1\right)\pmb{y}_1-\pmbmt{Q}_0\pmbmt{C}_1\pmb{y}_3\right].
\ee

\section{Toroidal Modes for poloidal magnetic fields}

Here, we revisit the problem of pure toroidal magnetic modes of a neutron star magnetized with a pure poloidal field,
corresponding to the case of $\zeta_0=0$.
The perturbed equations may be given as
\be
\omega^2\xi_\phi=-{1\over 4\pi \rho}\left(\pmb{B}\cdot\nabla B'_\phi+{B_\theta\cot\theta+B_r\over r}B'_\phi\right),
\ee
\be
B'_\phi=\pmb{B}\cdot\nabla\xi_\phi+\left({\partial B_r\over\partial r} +{B_\theta\cot\theta+B_r\over r}\right)\xi_\phi,
\ee
where $\pmb{B}=B_r\pmb{e}_r+B_\theta\pmb{e}_\theta$, and substituting the series expansions in terms of $Y_l^m$, we obtain for the expansion coefficients
the oscillation equations for a fluid star, which are given by
\begin{eqnarray}
r{{\rm d}\pmb{y}_5\over {\rm d}r}
=-{{\cal Q}\over f}\left({1\over 2}\pmbmt{B}_0^{-1}\pmbmt{W}_1\pmbmt{\Lambda}_0-\pmbmt{1}\right)\pmb{y}_5
+{1\over 2}\pmbmt{B}_0^{-1}\pmbmt{\Lambda}_1\pmb{y}_6,
\label{eq:y5t}
\end{eqnarray}
\begin{eqnarray}
r{{\rm d} \pmb{y}_6\over {\rm d}r}
=-{1\over 2\beta}c_1\bar\omega^2\pmbmt{B}_1^{-1}\pmbmt{\Lambda}_0\pmb{y}_5
-\left({1\over 2}{{\cal Q}\over f}\pmbmt{B}_1^{-1}\pmbmt{W}_0\pmbmt{\Lambda}_1+{d\ln fr\over d\ln r}\pmbmt{1}\right)\pmb{y}_6,
\label{eq:y6t}
\end{eqnarray}
and are solved for the frequency $\omega$ with appropriate boundary conditions at the centre and the surface of the star.
The inner boundary conditions are the regularity condition for the functions $\pmb{y}_5$ and $\pmb{y}_6$ at the center.
For the surface boundary condition, we may use $B'_\phi=0$, which is equivalent to the condition employed in this paper, or $\partial (\xi_\phi/r)/\partial r=0$, which could be applied at the surface of the solid crust of a neutron star so that the traction vanishes.
These two surface boundary conditions are represented by $\pmb{y}_6=0$ and ${{\rm d}\pmb{y}_5/{\rm d}r}=0$, respectively.
In Table C1, we tabulate $\bar\omega$ of pure toroidal magnetic modes for the two different surface boundary conditions.
For the condition $B'_\phi=0$, we find pure toroidal magnetic modes only for odd parity.
On the other hand, pure toroidal magnetic modes of even and odd parity
are found for the surface boundary condition $\partial (\xi_\phi/r)/\partial r=0$.
For this boundary condition, we find a complex eigen-mode having no radial nodes of the eigenfunctions of dominant amplitudes, which is the only complex mode we find.
The complex frequency, however, does not necessarily well converge with increasing $j_{\rm max}$, although 
the frequency of the real modes shows good convergence.
Wave patterns $\hat\xi_\phi$ of the pure toroidal modes calculated for the condition $B'_\phi$
are quite the same as those shown in Fig. \ref{fig:yievenz001b15_toroid}.
Wave patterns obtained for the condition $\partial (\xi_\phi/r)/\partial r=0$, however,
are quite different as shown by
Fig. \ref{fig:yievenz1b15_toroid}, in which the wave patterns $\hat\xi_\phi$ and $\hat B'_\phi$ of the pure toroidal magnetic mode
 of odd parity are depicted for $B_S=10^{15}$G, where $\bar\omega=1.923\times10^{-2}$.
The patterns we obtain for $\partial (\xi_\phi/r)/\partial r=0$ are well converged for $j_{\rm max}=14$ and
look similar to one of the patterns obtained by Cerd\'a-Dur\'an et al (2009).

We may understand a reason for the existence of complex eigenfrequencies using the energy equation derived in \S 2.3.
For pure toroidal modes, we have 
\be
\omega^2={1\over 4\pi}{\displaystyle \int dV \left|B'_\phi\right|^2-\int dS\xi_\phi^*B'_\phi B_r\over
\displaystyle \int dV\rho\left|\xi_\phi\right|^2}.
\label{eq:inteqtoroid}
\ee
Note that $\pmb{n}\cdot\pmb{B}\not=0$ at the surface of the star.
Equation (\ref{eq:inteqtoroid}) suggests that 
$\omega^2$ is always a real number for the surface boundary condition $B'_\phi=0$.
For $\partial (\xi_\phi/r)/\partial r=0$, on the other hand, the surface term becomes
\be
\int dS B_r\xi_\phi^*B'_\phi=\int dS B_r\left[B_\theta{1\over r}\xi_\phi^*{\partial\xi_\phi\over\partial\theta}
+\left|\xi_\phi\right|^2\left({1\over r}{\partial rB_r\over \partial r}+{B_\theta\over r}\cot\theta\right)\right],
\ee
which can be a complex number, leading to complex $\omega^2$.
Note that both $\omega$ and its complex conjugate $\omega^*$ can be a solution.
For the boundary condition $\partial(\xi_\phi/r)/\partial r=0$, the force operator $\pmb{F}(\pmb{\xi})$
cannot be self-adjoint, and hence complex solutions are permitted.

\begin{table*}
\begin{center}
\caption{Eigenfrequency $\bar{\omega}$ of pure toroidal magnetic modes for $B_{\rm S}=10^{15}$ G
for two different surface boundary conditions, where the largest eigenfrequency for a given number of radial nodes
is tabulated.}
\begin{tabular}{cccc}
\hline
\multicolumn{4}{c}{  $B'_\phi=0$ } \\
\hline
parity & \multicolumn{3}{c}{  number of radial nodes } \\
\hline
 & 0 & 1 & 2 \\
\hline
odd & 5.862$\times10^{-3}$ & 1.918$\times10^{-2}$  & 3.176$\times10^{-2}$\\
even & $\cdots$ & $\cdots$  & $\cdots$\\
\hline
\multicolumn{4}{c}{  ${\partial (\xi_\phi/r)/\partial r}=0$ } \\
\hline
parity & \multicolumn{3}{c}{  number of radial nodes } \\
\hline
 & 0 & 1 & 2 \\
\hline
  odd     & $\cdots~^a$ 
  & 1.923$\times10^{-2}$ & 3.186$\times10^{-2}$ \\
  even       &     $\cdots$      & $\cdots$ & 2.556$\times10^{-2}$ \\
\hline
\multicolumn{4}{l}{ $^a$ We obtain $\bar\omega=7.188\times10^{-3}\pm2.73\times10^{-4}\rmi$ for $j_{\rm max}=14$.}
\end{tabular}
\end{center}
\end{table*}

\begin{figure}
\begin{center}
\resizebox{0.4\columnwidth}{!}{
\includegraphics{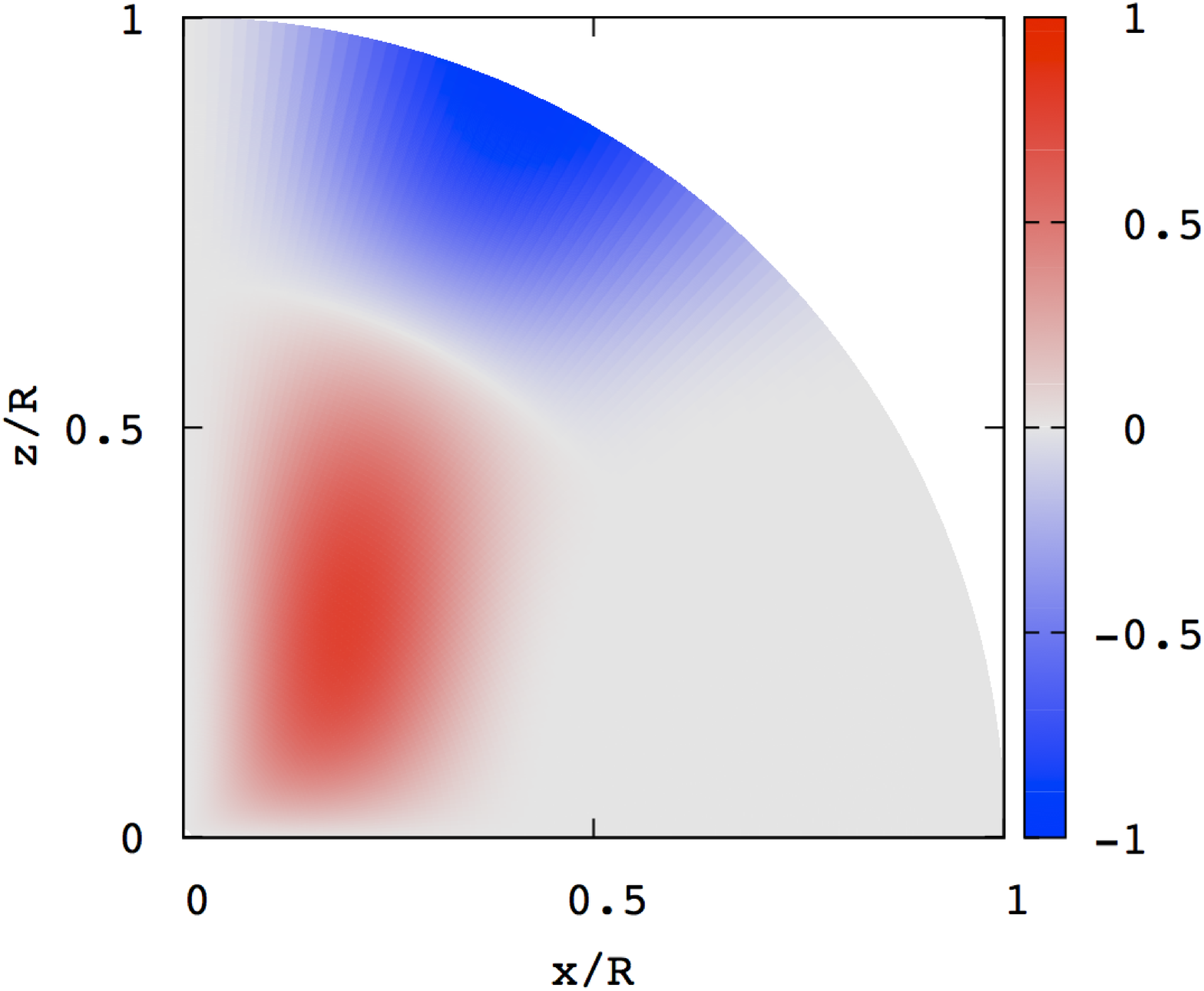}}
\resizebox{0.4\columnwidth}{!}{
\includegraphics{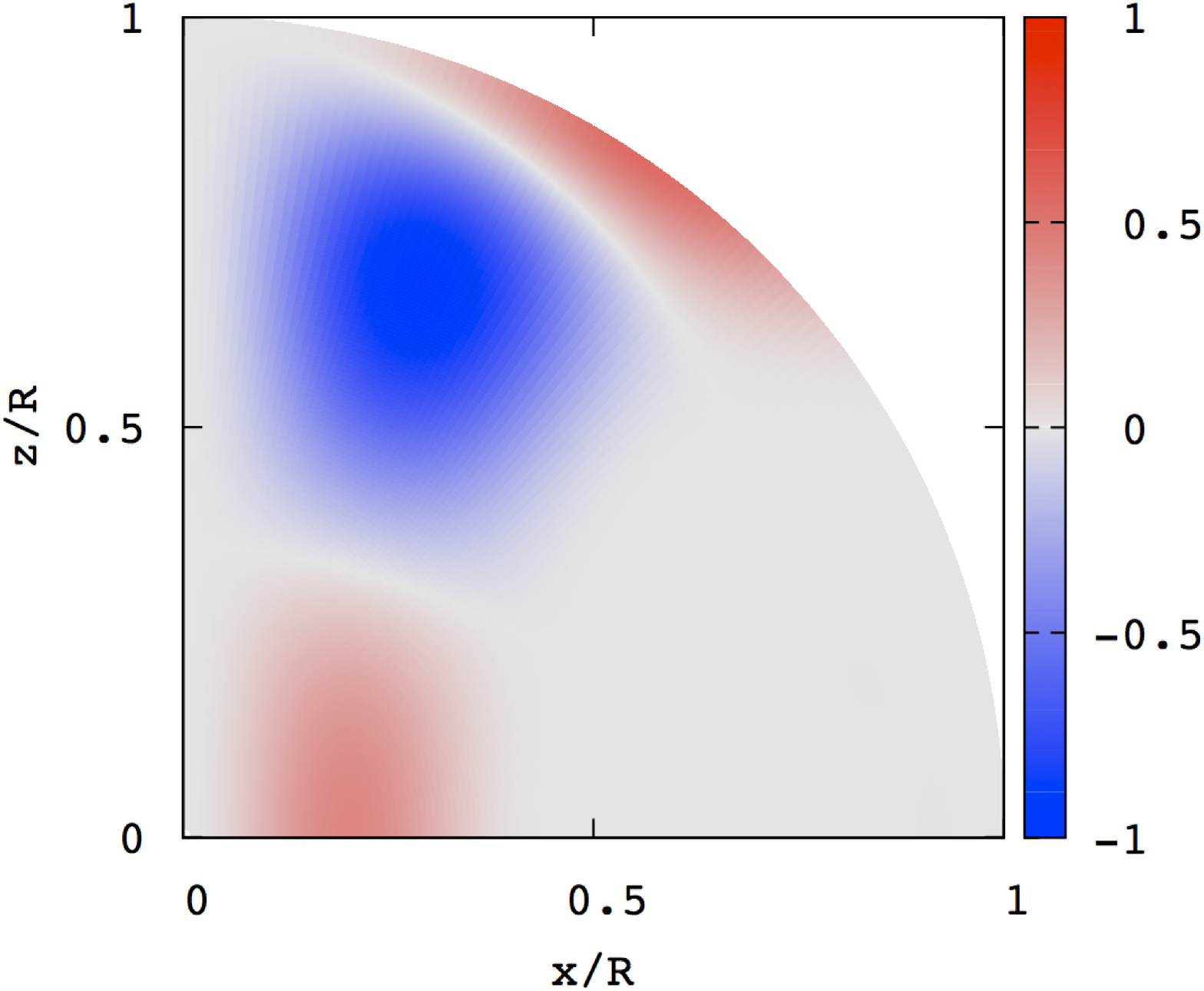}}
\end{center}
\caption{Wave patterns $\hat\xi_\phi$ (left) and $\hat B'_\phi$ (right) of the pure toroidal magnetic mode
of odd parity for $B_S=10^{15}$G where $\bar\omega=1.923\times10^{-2}$.
}
\label{fig:yievenz1b15_toroid}
\end{figure}

It is interesting to note that, using equations (\ref{eq:y5t}) and (\ref{eq:y6t}), we may derive
\be
{d^2\xi_\phi\over d\tau^2}+{d\ln\sqrt{4\pi\rho}\over d\tau}{d\xi_\phi\over d\tau}+\left(\omega^2-\Psi
\right)\xi_\phi=0,
\label{eq:xiphidtau}
\ee
where
\be
{d\over d\tau}={\pmb{B}\over\sqrt{4\pi\rho}}\cdot\nabla, \quad \Psi=\omega_A^2+{1\over\sqrt{4\pi\rho}}{d\over d\tau}\sqrt{4\pi\rho}\omega_A, \quad
\omega_A=-{1\over\sqrt{4\pi \rho}}{df\over dr}\cos\theta.
\ee
Equation (\ref{eq:xiphidtau}) may describe wave propagation along a magnetic field line, and 
the propagation region along a field line may be given by $\omega^2\ge\Psi$.
Wave propagation along a magnetic field line is not independent of that along different field lines
since the functions $\rho$ and $\Psi$ cannot be regarded as a function which depends only on $\tau$.
Ignoring the terms $d\xi_\phi/d\tau$ and $\Psi$, we obtain the wave equation similar to that used, for example, by Cerd\'a-Dur\'an et al
(2009).

\end{document}